\documentclass[twoside,english,american,noreviewchanges,noreferee,nodrafting,latin9,twocolappendix,apj,numberedappendix]{emulateapj}
\usepackage[T1]{fontenc}
\usepackage[latin9]{inputenc}
\usepackage{babel}
\usepackage{array}
\usepackage{textcomp}
\usepackage{url}
\usepackage{amstext}
\usepackage{amssymb}
\usepackage{graphicx}
\usepackage{esint}

\makeatletter

\let\SF@@footnote\footnote
\def\footnote{\ifx\protect\@typeset@protect
    \expandafter\SF@@footnote
  \else
    \expandafter\SF@gobble@opt
  \fi
}
\expandafter\def\csname SF@gobble@opt \endcsname{\@ifnextchar[
  \SF@gobble@twobracket
  \@gobble
}
\edef\SF@gobble@opt{\noexpand\protect
  \expandafter\noexpand\csname SF@gobble@opt \endcsname}
\def\SF@gobble@twobracket[#1]#2{}
\providecommand{\tabularnewline}{\\}

\bibpunct{(}{)}{;}{a}{}{,}
\usepackage{tabu}
\usepackage{threeparttable}
\usepackage{booktabs}
\usepackage{algorithm2e}
\newcommand{\valuetotalobscspacedensitylowzfrac}{{75}^{+4}_{-4}\%}
\newcommand{\valuetotalobscspacedensityaveragefrac}{{77}^{+4}_{-5}\%}

\newcommand{\valuetotalobscspacedensitypeakzfrac}{{83}^{+3}_{-3}\%}

\newcommand{\valuetotalobsclumdensityaveragefrac}{{74}^{+4}_{-5}\%}

\newcommand{\valueCTspacedensitylowzfrac}{{39}^{+7}_{-6}\%}
\newcommand{\valueCTspacedensityaveragefrac}{{38}^{+8}_{-7}\%}

\newcommand{\valueCTspacedensitypeakzfrac}{{46}^{+6}_{-5}\%}

\newcommand{\valueCTlumdensityaveragefrac}{{40}^{+6}_{-6}\%}

\newcommand{\change}[2]{#2}
\newcommand{\changed}[1]{#1}

\makeatother

\usepackage{listings}
\addto\captionsamerican{}
\addto\captionsenglish{}

\begin{document}

\title{Obscuration-dependent evolution of Active Galactic Nuclei}

\newcommand{\MPE}{Max Planck Institut f\"ur Extraterrestrische Physik Giessenbachstrasse, 85748 Garching, Germany}
\newcommand{\IMP}{Astrophysics Group, Imperial College London, Blackett Laboratory, Prince Consort Road, London SW7 2AZ, UK}
\newcommand{\China}{National Astronomical Observatories, Chinese Academy of Sciences, 100012, Beijing, China}
\newcommand{\Alamos}{Los Alamos National Laboratory, Los Alamos, NM, USA}
\newcommand{\Kentucky}{Department of Physics and Astronomy, University of Kentucky, 505 Rose Street, Lexington, Kentucky 40506, USA}
\newcommand{\Athens}{National Observatory of Athens, V.  Paulou  \& I.  Metaxa, 11532, Greece}
\newcommand{\Durham}{Department of Physics, Durham University, Durham DH1 3LE, UK}
\newcommand{\Philly}{Department of Physics, Drexel University, 3141 Chestnut Street, Philadelphia, PA 19104, USA}
\newcommand{\Irvine}{Center for Galaxy Evolution, Department of Physics and Astronomy, University of California, Irvine, 4129 Frederick Reines Hall Irvine, CA 92697, USA}
\newcommand{\Pasadena}{Cahill Center for Astrophysics, California Institute of Technology, 1216 East California Boulevard, Pasadena, CA 91125, USA.}

\shortauthors{Buchner et al.}
\altaffiltext{$\bigstar$}{\email{johannes.buchner.acad@gmx.com}}
\altaffiltext{1}{\MPE}
\altaffiltext{2}{\Athens}
\altaffiltext{3}{\Pasadena}
\altaffiltext{4}{\China}
\altaffiltext{5}{\IMP}
\altaffiltext{6}{\Durham}
\altaffiltext{7}{\Philly}


{}

\author{Johannes Buchner\altaffilmark{1,$\bigstar$}}

\author{Antonis Georgakakis\altaffilmark{1,2}}

\author{Kirpal Nandra\altaffilmark{1}}

\author{Murray Brightman\altaffilmark{1,3}}

\author{Marie-Luise Menzel\altaffilmark{1}}

\author{Zhu Liu\altaffilmark{4}}

\author{Li-Ting Hsu\altaffilmark{1}}

\author{Mara Salvato\altaffilmark{1}}

\author{Cyprian Rangel\altaffilmark{5}}

\author{James Aird\altaffilmark{6}}

\author{Andrea Merloni\altaffilmark{1}}

\author{Nicholas Ross\altaffilmark{7}}
\begin{abstract}
We aim to constrain the evolution of AGN as a function of obscuration
using an X-ray selected sample of $\sim2000$ AGN from a multi-tiered
survey including the CDFS, AEGIS-XD, COSMOS and XMM-XXL fields. The
spectra of individual X-ray sources are analysed using a Bayesian
methodology with a physically realistic model to infer the posterior
distribution of the hydrogen column density and intrinsic X-ray luminosity.
We develop a novel non-parametric method which allows us to robustly
infer the distribution of the AGN population in X-ray luminosity,
redshift and obscuring column density, relying only on minimal smoothness
assumptions. Our analysis properly incorporates uncertainties from
low count spectra, photometric redshift measurements, association
incompleteness and the limited sample size. We find that obscured
AGN with $N_{H}>{\rm 10^{22}\, cm^{-2}}$ account for $\valuetotalobscspacedensityaveragefrac$
of the number density and luminosity density of the accretion SMBH
population with $L_{{\rm X}}>10^{43}\text{ erg/s}$, averaged over
cosmic time. Compton-thick AGN account for approximately half the
number and luminosity density of the obscured population, and $\valueCTspacedensityaveragefrac$
of the total. We also find evidence that the evolution is obscuration-dependent,
with the strongest evolution around $N_{H}\thickapprox10^{23}\text{ cm}^{-2}$.
We highlight this by measuring the obscured fraction in Compton-thin
AGN, which increases towards $z\sim3$, where it is $25\%$ higher
than the local value. In contrast the fraction of Compton-thick AGN
is consistent with being constant at $\approx35\%$, independent of
redshift and accretion luminosity. We discuss our findings in the
context of existing models and conclude that the observed evolution
is to first order a side-effect of anti-hierarchical growth.
\end{abstract}
\maketitle

\section{Introduction}


Supermassive Black Holes (SMBH) are abundant in the local universe
-- every nearby massive galaxy harbours one \citep{Richstone1998,Kormendy2013}.
The majority of the growth of these SMBH must have been through accretion
processes \citep{Soltan1982,Merloni2008}. The material accreted by
the SMBH likely originates from galactic scales and therefore one
might expect some relation between the properties of galaxies and
their central black holes. This idea was reinforced by the discovery
of tight relationships between SMBH and the properties of the stellar
spheroid of galaxies \citep{Magorrian1998,Ferrarese2000,Gebhardt2000,Tremaine2002,Ferrarese2005,Kormendy2013},
which can be interpreted as evidence for co-evolution of AGN and their
host galaxies. For instance, the growth of AGN through accretion may
be linked to the host galaxies star formation, either weakly through
a shared gas reservoir, or strongly via so-called ``feedback'' processes.
In the latter, the energetic output from the AGN into the galactic
environment can strongly disturb the galaxy \citep{Silk1998,Fabian1999a}.

To understand the growth of SMBH over cosmic time and the relationship
of that growth to galaxy evolution, it is first necessary to estimate
the space density of accreting SMBH, and its evolution over cosmic
time in an unbiased way. This is challenging because it is known that
many AGN are enshrouded in gas and dust, making them difficult to
detect directly. X-ray emission is an efficient way to reveal these
obscured SMBH, at least for moderate column densities. On the other
hand when the column of equivalent neutral hydrogen exceeds unit optical
depth corresponding to the Thomson cross section ($N_{H}\approx1.5\times10^{24}\,\text{ cm}^{-2}$)
AGN become difficult to find even at X-ray wavelengths. In this so-called
``Compton-thick'' regime the AGN can none the less still be identified
by hard X-ray emission which can emerge from the thick covering and/or
via radiation that is reflected and scattered into the line of sight.
These processes result in a characteristic shape to the X-ray spectrum,
including a flat continuum and intense iron K$\alpha$ emission lines
which can be identified using spectral analysis. The contribution
of these most heavily obscured AGN represents the major remaining
uncertainty in our knowledge of the accretion history, and placing
limits on their contribution is therefore vital.

Previous work has attempted to constrain the number of Compton-thick
AGN using X-ray background synthesis \citep[e.g.][]{Gilli2007}, but
while this population is needed to reproduce the shape of the XRB
spectrum, the method is relatively insensitive to the precise Compton-thick
AGN fraction \citep{Akylas2012}. Multi-wavelength data have also
been exploited to identify Compton-thick AGN and constrain their number.
These include optical \changed{\citep{Risaliti1999,Cappi2006,Panessa2006,Akylas2009,Gilli2010,Vignali2010,Jia2013,Mignoli2013,Vignali2014}},
infrared \changed{\citep[e.g.][]{Fiore2008,Fiore2009,Alexander2011,Brightman2011a,Brightman2011b}}
and hard X-ray \changed{\citep[e.g.][]{Sazonov2008,Burlon2011,Alexander2013,Lanzuisi2014}}
diagnostics applied to local\change{}{ and non-local} AGN samples.
These studies estimate Compton-thick fractions relative to the overall
obscured AGN population in the range $30-50\%$, thereby demonstrating
that such heavily obscured sources represent a sizeable fraction of
the AGN population in the nearby Universe. Following a handful of
early discoveries of Compton-thick AGN in deep X-ray surveys \citep{Norman2002,Tozzi2006,Comastri2011},
an important recent development has been the identification \changed{}{of}
significant samples of Compton-thick AGN at moderate to high redshifts
\citep{Brightman2012,Georgantopoulos2013,Buchner2014,Brightman2014}.
This has been enabled by the combination of extremely deep X-ray data,
sufficient to constrain the X-ray spectra, along with extensive multi-wavelength
coverage of X-ray survey regions, and new techniques able to determine
accurate photometric redshifts for X-ray emitting AGN \citep{Salvato2009,Salvato2011}.
This then offers the exciting possibility of starting to constrain
the evolution of the obscured AGN populations, including Compton-thick
AGN, provided that the selection function can be sufficiently well
understood.

Further interest in the obscured AGN population lies in the nature
of the obscuration itself, and its possible relationship with other
source properties. In the standard unification picture \citep{AntonucciUnification1985,Antonucci1993},
all AGN are surrounded by an optically thick toroidal structure relatively
close (parsec-scale) to the central engine, which can obscure the
line of sight depending on the viewing angle. Alternatively, or additionally,
obscuration can occur at galactic scales \citep{Maiolino1995}. Observations
show that AGN host galaxies are massive and lie, at least in an average
sense, on the main sequence of star-formation \citep{Santini2012}.
At moderate redshifts, $z\approx1-2$, such galaxies are known to
be gas-rich, with gas contents 3 to 10 times larger than local samples
\citep{Tacconi2013}. It is therefore possible that obscuration in
moderate redshift AGN is associated with the same gas fuelling both
star formation and the accretion process itself. In some scenarios
\citep{Hopkins2006OriginModel,Hopkins2011}, obscured AGN represent
a distinct phase in the co-evolution of the galaxy and its central
black hole, with energy output from accretion sweeping up gas from
the surroundings with potentially profound effects on star formation
\citep{Silk1998,Fabian1999,King2003}. Determining accurately the
fraction of obscured AGN, including Compton-thick AGN, as a function
of other parameters such as luminosity and redshift is critical for
both unification and co-evolution models. Previous work has provided
evidence that the obscured fraction depends on luminosity, with obscuration
being less common in more powerful sources \changed{\citep{Ueda2003,Treister2004,Akylas2006,Hasinger2008,Ebrero2009,Ueda2014}},
with local samples suggesting that the obscured fraction may peak
at around $L_{{\rm X}}\sim10^{42}\text{ erg/s}$ \citep{Brightman2011b,Burlon2011}.
More controversially, it has been suggested that the obscured fraction
increases out to high redshift (\changed{\citealp{LaFranca2005,Hasinger2008,Brightman2012,Iwasawa2012a,Vito2014}};
but see \citealp{Ueda2003,Akylas2006,Ebrero2009}). 

The same deep X-ray data and \change{multi wavelength}{multi-wavelength}
supporting data needed to identify Compton-thick AGN also enable an
accurate determination of the obscured AGN fraction and its evolution,
again with the proviso that the selection function must be well understood.
This is, however, very challenging, given the limited photon statistics
in deep X-ray data, and associated uncertainties in X-ray spectral
analysis, combined with additional uncertainties e.g. in photometric
redshifts. Moreover, studies of nearby objects show that the X-ray
spectra of AGN are complex, including multiple emission components,
e.g. absorption, reflection or scattering of direct AGN emission. 

In this paper we develop a novel non-parametric method for determining
the space density of AGN as a function of accretion luminosity, redshift
and hydrogen column density. We build on the X-ray spectral analysis
of \citet{Buchner2014}, applying their Bayesian spectral analysis
technique of a realistic, physically motivated model to a multi-layered
survey determine the luminosity and level of obscuration in a large
sample of X-ray selected AGN across a wide range of redshifts. Important
features of our approach are that, first, all sources of uncertainty
are consistently factored into the analysis and secondly, unlike previous
studies, no functional form is imposed. Instead, we use a smoothness
assumption to allow the data to determine the space density of AGN
and its dependence on luminosity, redshift and the level of obscuration
along the line of sight. The methodology is designed to be informative
at regions of the parameter space where data are sparse and therefore
constraints are expected to be loose. 

We adopt a cosmology of $H_{0}=70\,\text{km}\,\text{s}^{-1}\text{Mpc}^{\text{\textminus}1}$
, $\Omega_{M}=0.3$ and $\Omega_{\Lambda}=0.7$. Solar abundances
are assumed. The galactic photo-electric absorption along the line
of sight direction \change{to the CDFS is modelled with $N_{H}\approx8.8\times10^{19}\,\text{ cm}^{-2}$
\citep{Stark1992GalHImap}}{is modelled with $N_{H}\approx8.8\times10^{19}$,
$1.3\times10^{20}$, $2.7\times10^{20}$ and $2.2\times10^{21}\,\text{ cm}^{-2}$
for the CDFS, AEGIS-XD, COSMOS and XMM-XXL fields respectively \citep{Dickey1990,Stark1992GalHImap,Kalberla2005GalNHdist}}.
\change{The same value was used for the other survey fields. }{}In
this work, luminosity ($L$) always refers to the intrinsic (absorption-corrected)
luminosity in the $2-10\,\text{ keV}$ rest-frame energy range.

\section{\label{sec:data}Data}

The determination of the obscured fraction of AGN as a function of
redshift, accretion luminosity requires good coverage of the $L_{X}-z$
plane. We therefore combine X-ray survey fields with different characteristics
in terms of depth and areal coverage. These include the Chandra Deep
Field South \citep[CDFS,][]{Xue2011}, the All Wavelength Extended
Groth strip International Survey \citep[AEGIS,][]{Davis2007}, the
Cosmological evolution Survey \citep[COSMOS,][]{Scoville2007} and
the equatorial region of the XMM-XXL survey (PI: Pierre).

In our analysis we focus on the region of the CDFS which is covered
by the 4\,Ms Chandra observations and the part of the AEGIS field
which has been surveyed by Chandra to an exposure of 800\,ks (AEGIS-XD,
Nandra et al., submitted). In the COSMOS field we use the region covered
by the Chandra observations performed between November 2006 and June
2007 \citep[C-COSMOS,][]{Elvis2009}. Table \ref{tab:Survey-fields}
presents information on the individual X-ray fields used in this paper.

\begin{table*}[t]
\protect\caption{\label{tab:Survey-fields}Survey fields}

\centering
\begin{threeparttable}

\begin{tabular*}{0.99\textwidth}{@{\extracolsep{\fill}}ccccc}
\toprule[1.5pt]Survey & CDFS & AEGIS-XD & C-COSMOS & XMM-XXL\tabularnewline
\midrule
Survey area  & 464 arcmin$^{2}$ & 1010 arcmin$^{2}$  & 0.9 deg$^{2}$  & 20 deg$^{2}$ \tabularnewline
Total/central exposure time  & 4 Ms & 2.4 Ms/800 ks  & 1.8 Ms/160 ks & 10 ks\\
\bottomrule[1.5pt] 
\end{tabular*}

\end{threeparttable}
\end{table*}

\begin{table*}[ht]
\protect\caption{\label{tab:Steps}Sample statistics for the individual data extraction
steps.}

\centering
\begin{threeparttable}

\begin{tabular*}{0.99\textwidth}{@{\extracolsep{\fill}}>{\raggedright}p{3em}>{\raggedright}p{6cm}ccccc}
\toprule[1.5pt] Section & Extraction step & CDFS & AEGIS-XD & C-COSMOS & XMM-XXL & total\tnote{*}\tabularnewline
\midrule
2.1,2.2 & X-ray hard-band detected  & 326 & 574 & 1016 & 206 & 2122\tabularnewline
2.3  & Association with optical / IR  & 315 & \change{559}{559} & 1016 & 206 & \change{2096}{2096}\tabularnewline
2.4 & Redshift information: \\
\qquad{}spec-z/photo-z/no-z/removed stars  & 180/131/11/4 & 227/\change{336}{322}/\change{15}{15}/\change{10}{10} & 491/519/0/6 & 174/0/32/0 & 1072/986/\change{58}{58}/\change{20}{20}\tabularnewline
2.5 & X-ray spectral extraction and data analysis: \\
\qquad{}successful/failed extraction & 321/1 & \change{564}{564}/- & 1010/- & 206/- & \change{2101}{2101}/1\tabularnewline
2.6 & galaxies removed (based on \change{$L_{X}<42$}{$L_{2-10\text{ keV}}<10^{42}\text{ erg/s}$}) & 20 & \change{11}{11} & \change{15}{14} & 1 & \change{47}{46}\tabularnewline
2.7 & Objects used for LF analysis & 302 & \change{553}{553} & \change{995}{996} & 205 & \change{2055}{2056}\\ \bottomrule[1.5pt] 
\end{tabular*}

\begin{tablenotes}

\item[*] Total number of sources selected.

\end{tablenotes}

\end{threeparttable}
\end{table*}

\subsection{\label{sec:detect}\label{sec:detect_chandra}Chandra observations}

The data reduction, source detection and source catalogues for two
of the Chandra fields, the CDFS \citep{Rangel2013} and the AEGIS-XD
(Nandra et al., submitted), follow the methodology described by \citet{Laird2009}.
Briefly, hot pixels, cosmic ray afterglows and times of anomalously
high backgrounds are removed to produce clean level-2 event files.
These are then aligned using bright sources and subsequently merged.
Images and exposure maps are constructed in four energy bands, $0.5-2$,
$2-7$, $5-7$ and $0.5-7$\,keV. A candidate source list is created
using \texttt{wavdetect} at a low significance threshold ($10^{-4}$).
Source and background counts are then extracted at each candidate
source position. The source region corresponds to the $70\%$ encircled
energy fraction (EEF) of the point spread function (PSF). The background
count region is an annulus with inner radius 1.5 times the 90\% EEF
of the PSF and width 100 pixel of 0.5\,arcsec in size. For the determination
of the background, candidate sources are masked out. For each candidate
source position, the Poisson probability that the observed counts
are a background fluctuation is computed. Sources are accepted if
that probability is $<4\times10^{-6}$ (\citealp{NandraSrcDet2005}).
In this paper we use the hard band ($2-7$\,keV) selected sample.
The X-ray sensitivity curves are estimated by extrapolating the background
counts and exposure maps in the 2-7\,keV band to the limiting flux
of a source in the 0.5-10\,keV energy range by adopting the methods
described in \citet{Georgakakis2008}. For the C-COSMOS survey we
use the \change{2-10}{2-7}\,keV selected X-ray source catalogue
presented by \citet{Elvis2009}. The number of X-ray sources detected
are broken down by field in Table \ref{tab:Steps}. The corresponding
sensitivity curves are shown in Figure \ref{fig:Area-curve}.

\subsection{\label{sec:detect_xmm}XMM observations}

The Chandra surveys above need to be complemented by a shallower and
wider X-ray field to place constraints at the bright end of the luminosity
function. For this we use the XMM-XXL survey, which consists of ${\rm 10\, ks}$
XMM pointings that cover a total area of ${\rm 50\, deg^{2}}$ split
into two equal size fields. In this paper we focus on the equatorial
sub-region of the XMM-XXL. The data reduction, source detection and
sensitivity map construction follow the methods described in \citet{Georgakakis2011}.
A full description of the XMM-XXL X-ray source catalogue generation
are presented by Liu et al. (in prep.). The most salient details of
those steps are outlined here.

The XMM observations were reduced using the Science Analysis System
(SAS) version 12. The first step is to produce event files from the
Observation Data Files (ODF) using the \textsc{epchain} and \textsc{emchain}
tasks of \textsc{sas} for the EPIC \citep[European Photon Imaging Camera;][]{Struder2001EPIC,Turner2001EPIC}
PN and MOS detectors respectively. Pixels along the edges of the CCDs
of the PN and MOS detectors are removed because their inclusion often
results to spurious detections. Flaring background periods are identified
and excluded using a methodology similar to that described in \citet{Nandra2007}.
Images and exposure maps in celestial coordinates with pixel size
of 4.35\,arcsec are constructed in 5 energy bands, $0.5-8$, $0.5-2$,
$2-8$, $5-8$ and $7.5-12$\,keV. All overlapping EPIC images are
merged prior to source detection to increase the sensivity to point
sources. The detection algorithm is applied independently to each
of the 5 spectral bands defined above.

The source detection methodology is similar to that described in \citet{Laird2009}
in the case of Chandra data. Source candidates are identified using
the wavelet-based \textsc{ewavelet} source detection task of \textsc{sas}
at a low threshold of $4\sigma$ above the background, where $\sigma$
is the RMS of the background counts. For each candidate source the
Poisson probability of a random background fluctuation is estimated.
\changed{This step involves the extraction of the total counts at
the position of the source and the determination of the local background
value. To match the asymmetric PSF of XMM, especially off-axis, elliptical
apertures were used from the XMM EPIC PSF parametrisation of \citet{Georgakakis2011}.
The count extraction region is obtained by scaling the elliptical
apertures to contain 70 per cent of the PSF EEF. The total counts
at a candidate source position, $T$, is the sum of the extracted
counts from individual EPIC cameras. For each source the local background
is estimated by first masking out all detections within 4\,arcmin
of the source position using an elliptical aperture that corresponds
to the 80 per cent EEF ellipse. The counts from individual EPIC cameras
are then extracted using elliptical annuli centred on the source with
inner and outer semi-major axes of 5 and 15 pixels (0.36 and 1.09\,arcmin)
respectively, while keeping the same shape as the elliptical aperture
in terms of rotation and ellipticity.} The mean local background,
$B$, is then estimated by summing up the background counts from individual
EPIC cameras after scaling them down to the area of the source count
extraction region. The Poisson probability $P(T,\, B)$ that the extracted
counts at the source position, $T$, are a random fluctuation of the
background is calculated. We consider as sources the detections with
$P(T,\, B)<4\times10^{-6}$. The above methodology is optimised for
the detection of point sources. The final catalogue however, includes
extended X-ray sources associated with hot gas from galaxy clusters
or groups. Also, the extended X-ray emission regions of bright clusters
are often split into multiple spurious detections by our source detection
pipeline. The \textsc{emldetect} task of \textsc{sas} is used to identify
extended sources (i.e. groups or clusters) and spurious detections.
Point sources for which \textsc{emldetect} failed to determine a reliable
fit and hence are considered spurious and excluded from the analysis.
The \textsc{eposcorr} task of \textsc{sas} is used to correct for
systematic errors in the astrometric positions of X-ray sources by
cross-correlating with optically sources in the SDSS-DR8 catalogue
\citep{Aihara2011SDSSIII} with magnitudes $r<22$\,mag.

The flux of each source in different spectral bands is estimated by
assuming a power-law X-ray spectrum with $\Gamma=1.4$, i.e. similar
to the XRB, absorbed by the appropriate Galactic hydrogen column density.
The latter is derived from the HI map of \citet{Kalberla2005GalNHdist}
using the right ascension and declination of the aimpoint of each
XMM observation and the \textsc{nh} task of \textsc{ftools}. The energy
to flux conversion factors are such that the counts from the 0.5-2,
0.5-8, 2-8, 5-8 and 7.5-12\,keV bands are transformed to fluxes in
the 0.5-2, 0.5-10, 2-10, 5-10 and 7.5-12\,keV bands respectively.

In this work we use the XMM-XXL point source sub-sample that is selected
in the 2-8\,keV band. We minimise optical identification and spectroscopic
redshift determination incompleteness (see next sections) by applying
a bright flux cut, $f_{X}({\rm 2-10\, keV)>7\times10^{-14}\, erg\, s^{-1}\, cm^{-2}}$.
The total number of XMM-XXL X-ray sources used in the analysis is
206 (see Table \ref{tab:Steps}).

The construction of the sensitivity maps follows the methodology of
\citet{Georgakakis2008}. The 2-10\,keV sensitivity curve of the
sample is plotted in Figure \ref{fig:Area-curve}. The window function
of the spectroscopic follow-up observations in the XMM-XXL field (see
Section \ref{sec:redshifts}) is taken into account in this calculation.
Also, in addition to the Poisson false detection probability threshold
of $<4\times10^{-6}$ in the sensitivity map calculation we also take
into account the flux cut $f_{X}({\rm 2-10\, keV)>7\times10^{-14}\, erg\, s^{-1}\, cm^{-2}}$,
by estimating at each survey position the probability of measuring
a flux above this limit.

The number of X-ray sources detected -- about 2000 in total -- are
broken down by field in Table \ref{tab:Steps}. The corresponding
sensitivity curves are shown in Figure \ref{fig:Area-curve}. We associate
these detected X-ray source positions to optical counterparts in order
to obtain redshifts.

\subsection{\label{sec:association}Association with optical/IR counterparts}

The identification of the X-ray sources with optical or infrared counterparts
in the AEGIS-XD, COSMOS and XMM-XXL used the Likelihood Ratio method
of \citet{SutherlandSaunders1992}. \change{In the case of the AEGIS-XD
the identifications are described in}{Specific details on the association
of X-ray sources with optical/infrared counterparts are presented
by} Nandra et al. (submitted). They used the multi-waveband photometric
catalogue provided by the Rainbow Cosmological Surveys Database \citep{Perez-Gonzalez2008,Barro2011a,Barro2011b}.

The counterparts of C-COSMOS X-ray sources are taken from \citet{Civano2012}.
For the identification they used the $I$-band selected optical sample
of \citet{Capak2007}, and the $K$-band photometry of \citet{McCracken2010}
and the IRAC-$\mathrm{3}.6\mu m$ catalogue of \citet{Sanders2007}.

X-ray sources in the XMM-XXL survey were matched to the SDSS-DR8 photometric
catalogue \citet{Aihara2011SDSSIII} following the methods described
in \citet{Georgakakis2011}.

\citet{HsuCDFSz2013} presents the counterparts of the CDFS X-ray
sources. They used a sophisticated Bayesian version of the Likelihood
Ratio method which is based on the probabilistic formalism of \citet{Budavari2008}.
The photometric catalogues used to identify X-ray sources include
the CANDELS $H$-band selected multi-wavelength catalog in \citet{Guo2013},
the MUSYC catalogue presented by \citet{Cardamone2010} and the TENIS
near-infrared selected source catalogue described by \citet{Hsieh2012}.

\subsection{\label{sec:redshifts}Redshift estimation}

The X-ray survey fields used in this paper benefit from extensive
spectroscopic campaigns that also target specifically X-ray sources.
In the CDFS, we used the spectroscopic redshifts compiled by N. Hathi
(private communication, see \citealp{HsuCDFSz2013}). Spectroscopic
redshift measurements of X-ray sources in the AEGIS field are extracted
from the compilation presented in Nandra et al. (submitted) which
included also the DEEP2 \citep{Newman2012} and DEEP3 galaxy redshift
surveys \citep{Cooper2011DEEP3,Cooper2012DEEP3} as well as observations
carried out at the MMT using the Hectospec fibre spectrograph \citep{Coil2009}.
Redshifts in the C-COSMOS are used from the compilation of \citet{Civano2012}
which includes the public releases of the VIMOS/zCOSMOS bright project
\citep{Lilly2009zCOSMOS} and the Magellan/IMACS observation campaigns
\citep{Trump2009XMMCOSMOS}.

In the case of XMM-XXL, optical spectroscopy is from \citet{Stalin2010zXMMXXL},
the Baryon Oscillation Spectroscopic Survey \citep[BOSS;][]{Dawson2013BOSS,Bolton2012,Smee2013SDSS}
as well as dedicated Sloan Digital Sky Survey III (SDSS: \citealp{York2000SDSS,Gunn2006SDSS};
SDSS-III: \citealp{Eisenstein2011SDSS}) ancillary science observations,
which targeted specifically X-ray sources in the equatorial XMM-XXL
field (PI: A. Merloni, A. Georgakakis). Targets were selected to have
$f_{X}({\rm 0.5-10\, keV)>10^{-14}\, erg\, s^{-1}\, cm^{-2}}$ and
$17<r<22.5$, where $r$ corresponds either the PSF magnitude in the
case of optical unresolved sources (SDSS type=6) or the model magnitude
for resolved sources (see Menzel et al., in prep.). At the flux limit
$f_{X}({\rm 2-10\, keV)>7\times10^{-14}\, erg\, s^{-1}\, cm^{-2}}$,
84\% (174/207) of the XMM-XXL sources have secure redshift measurements.

The CDFS, AEGIS-XD and COSMOS \change{and }{}fields have multiwavelength
photometric observations that allow photometric redshift estimates
for sources that lack spectroscopy. The methods developed by \citet{Salvato2009,Salvato2011}
are adopted to achieve photometric redshift accuracies for X-ray AGN
comparable to galaxy samples. The photometric redshifts catalogues
are presented in \citet{HsuCDFSz2013} for the CDFS, Nandra et al.
(submitted) for the AEGIS-XD\change{}{ field} and \citep{Salvato2011}
for the COSMOS\change{}{ field}. The photometric redshifts are included
in the analysis in the form of probability distribution functions,
incorporating systematic uncertainties (see below). A by-product of
the photometric redshift determination is the characterisation of
the Spectral Energy Distribution (SED) of X-ray sources. This information
is used to identify candidate Galactic stars in the X-ray sample and
exclude from the analysis. Following \citet{Salvato2011} if stellar
templates provide an improved fit to the SED of a source, as measured
by the reduced $\chi^{2}$ i.e. $\chi_{\text{star}}^{2}<\chi_{\text{gal}}^{2}/1.5$,
and the source is point-like in the optical images, then it is considered
to be a Galactic star candidate. The number of removed stars is indicated
in Table \ref{tab:Steps}.

For the XMM-XXL field, multiwavelength photometry from the UV to the
infrared which homogeneously covers the surveyed area is not available.
This is essential for reliable photometric redshifts, especially in
the case of bright X-ray samples like the XMM-XXL \citep[e.g.][]{Salvato2011}.
Therefore, for the number of X-ray sources without spectroscopic redshift
measurement in that field (32/206) we choose not to determine photometric
redshifts. These sources are still included in the analysis by assigning
them a flat redshift prior (see below for details). In the XMM-XXL
bright sub-sample used here, there are no stars based on inspection
of the images.

\begin{figure*}[t]
\begin{centering}
\includegraphics[width=0.99\textwidth]{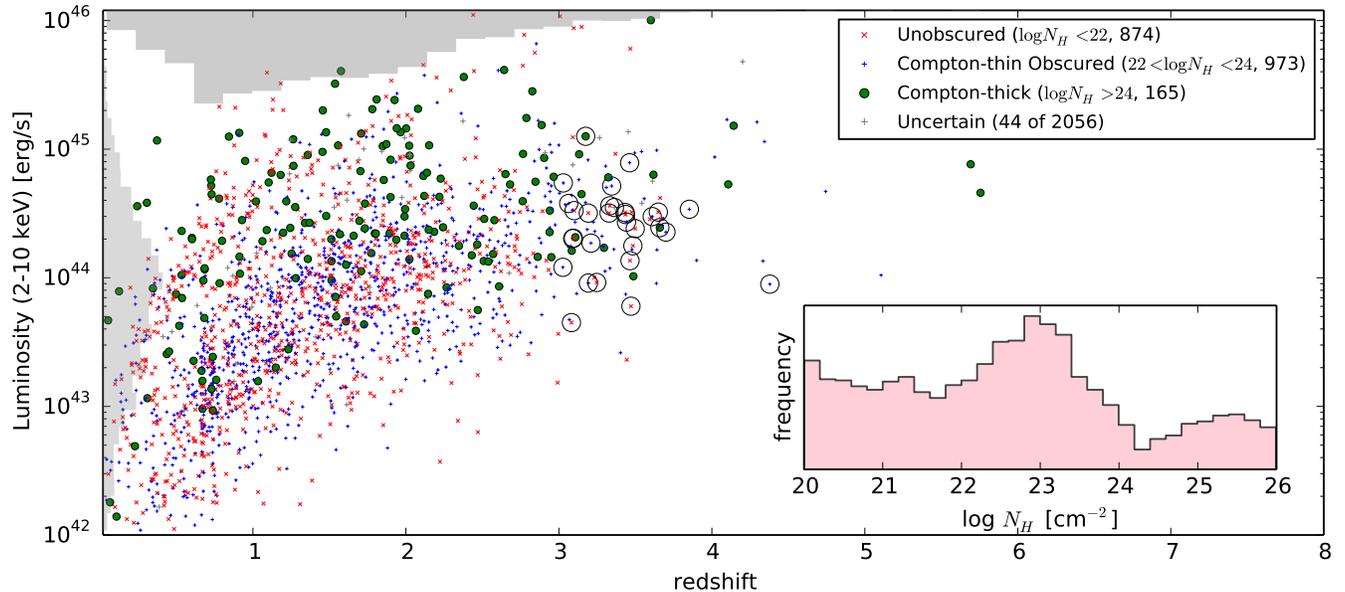}
\par\end{centering}

\protect\caption{\label{fig:Luminosity-redshift}Luminosity-redshift plot of the full
sample. The intrinsic $2-10\,\text{ keV}$ luminosity in $\text{ erg/s}$
and redshift is plotted. For this visualisation, we colour sources
based on $\log N_{H}$ (in units of $\text{ cm}^{-2}$ as follows:
$20-22$, ``Unobscured'', red; $22-24$, ``Compton-thin Obscured'',
blue; $24-26$, ``Compton-thick'', green). If more than $50\%$
of the $N_{H}$ posterior probability lies within one of the intervals
above the source is colour-coded accordingly. \change{The green points
are elevated to higher luminosities due to selection bias against
faint Compton-thick AGN. }{Due to the heavy suppression of the flux
of Compton thick objects by absorption, they are typically only detectable
at higher intrinsic luminosity compared to unobscured or Compton-thin
sources.} Some objects lie below the $L=10^{42}\text{ erg/s}$ limit
and are not used in this work. The inset shows the $N_{H}$ histogram
of our sample\change{}{, while the top and left axes show histograms
of the redshift and luminosity distributions in gray}. \change{This
is constructed by drawing a random posterior sample for each object.
The histogram thus includes the uncertainty in the parameter estimation
as well as the intrinsic distribution.}{ These plots are constructed
by drawing a random posterior sample for each object, and thus includes
the uncertainty in the parameter estimation as well as the intrinsic
distribution. Sources above redshift $z=3$ which have spectroscopic
redshift estimates are indicated with large black circles.}}
\end{figure*}

To summarise, the redshift determination falls in one of three cases
for each X-ray source: 
\begin{enumerate}
\item No redshift information available\change{}{ (58 sources)}: This
is the case if no spectroscopic redshifts are available, and photometric
redshift estimation is not possible due to limited photometry. This
case also applies when no secure association was found. Such a source
is associated with a flat redshift distribution in the interval $z=0.001-7$.
\item Spectroscopic redshifts\change{}{ (1072 sources)}: Wherever secure
spectroscopic redshifts are available, they are used directly, i.e.
do not associate uncertainty to them. \change{}{Different surveys
use different conventions to define the reliability of the redshift
measurement. In this paper we only consider spectroscopic redshifts
from the top two quality classes of any study, which typically corresponds
to a probability better than 95 per cent of being correct.}
\item Photometric redshifts\change{}{ (986 sources)}: These are included
in the analysis in the form of probability distribution functions.
We also incorporate systematic uncertainties e.g. due to incorrect
association to an optical counterpart (see Appendix B). 
\end{enumerate}
Table \ref{tab:Steps} shows a breakdown of how many sources fall
into each category. \change{}{The distribution of redshifts is plotted
in Figure \ref{fig:Luminosity-redshift}}.

\subsection{\label{sec:spectra}Extraction and analysis of X-ray spectra}

For the Chandra surveys, the \texttt{ACIS EXTRACT} (AE) software package
\citep{BroosACISEXTRACT2010} was used to extract spectra for each
source. The extraction followed the same methodology described in
\citet{Brightman2014}. Initially, each source and each pointing is
dealt with separately. AE simulates the PSFs at each source position.
Regions enclosing $90\%$ PSF at 1.5keV were used to extract source
spectra. The background regions are constructed around the sources
such that they contain at least 100 counts, with other sources masked
out. AE also constructs local response matrix files (RMF) and auxiliary
matrix files (ARF) using ray-tracing. As a final step, AE merges the
extracted spectra so that each source has a single source spectrum,
a single local background spectrum, ARF and RMF.

In the XMM-XXL field, X-ray source and background spectra were extracted
separately for each EPIC camera using \textsc{sas} version 12.0.1.
The source extraction regions were chosen to maximise the signal-to-noise
ratio using the \textsc{eregionanalyse} task of \textsc{sas}. In the
case of EPIC-PN camera, the background region has a radius of 90\,arcsec
and is placed, after visual inspection, on a source-free region on
the same CCD and raw-Y axis as the corresponding source. In the case
of EPIC-MOS because of the larger size of individual CCDs it is possible
to define the background region as an annulus with the inner and outer
radius were 40, 110\,arcsec respectively. If that was not possible,
because e.g. a source might lie close to CCD gaps then the background
regions was a circle of 90\,arcsec radius placed, on a source-free
region on the same CCD as the corresponding source. ARFs and RMFs
are generated using the \textsc{arfgen} and \textsc{rmfgen} tasks
of \textsc{sas}. Finally, the ARFs, RMFs, source and background spectra
of the same object observed by different EPIC cameras are merged by
weighting with the exposure time to produce unique spectral products
for each source.

\begin{figure}[h]
\begin{centering}
\includegraphics[width=0.5\textwidth]{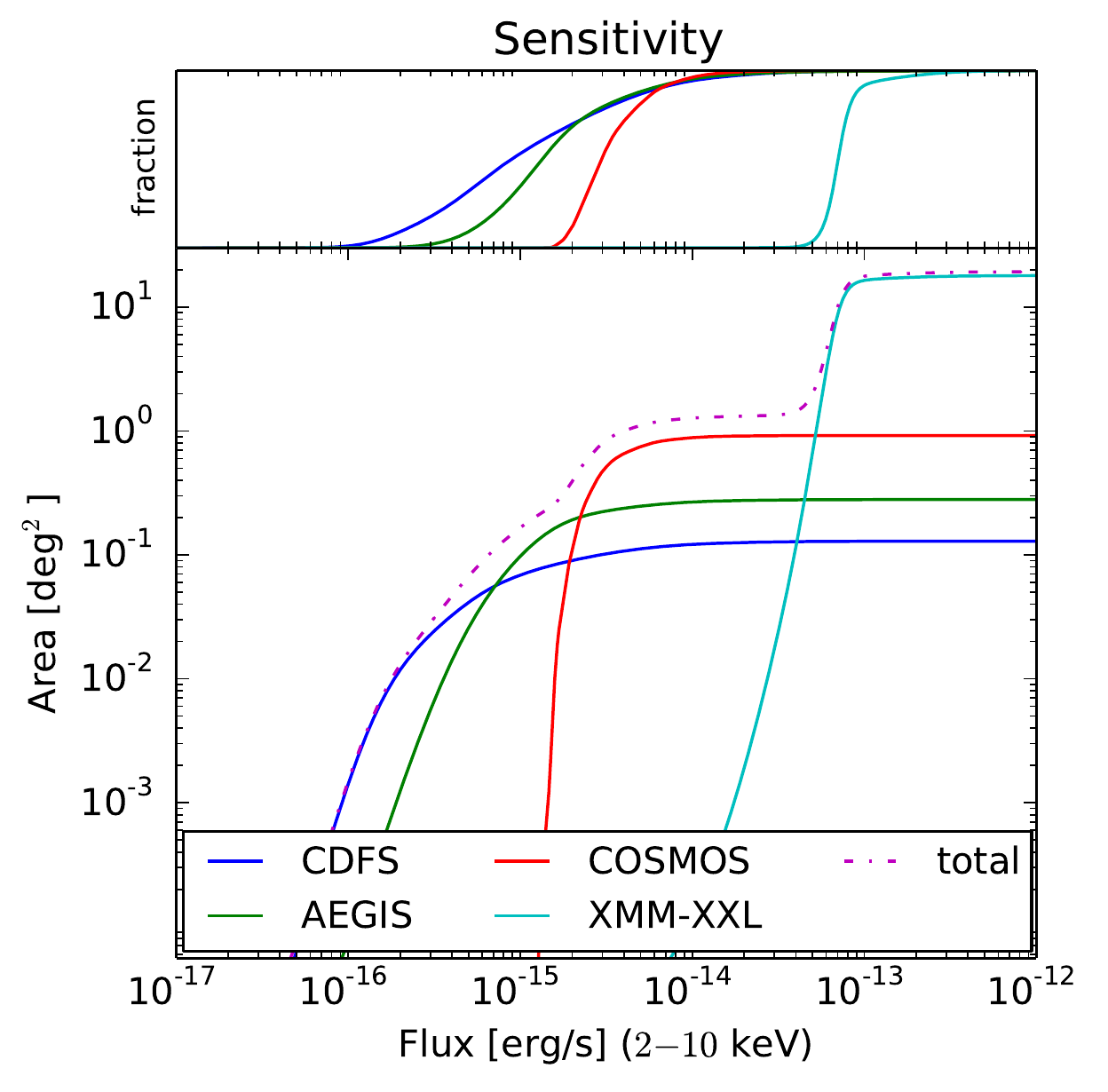}
\par\end{centering}

\protect\caption{\label{fig:Area-curve}Area curve. The \emph{bottom plot} shows all
X-ray sensitivity curves on a logarithmic scale for the sum of all
fields (dotted magenta line) and the individual fields. The \emph{top
plot} shows the sensitivity curves for the individual fields on a
linear scale and normalised to their respective maximum area (see
Table \ref{tab:Survey-fields}), to indicate the flux limit for detection.
The XMM-XXL curve is limited in flux by $7\times10^{-14}\,\text{ erg/s}$,
but we incorporate the uncertainty of measuring a higher flux due
to Poisson variance. This introduces a smooth transition.}
\end{figure}

Each X-ray spectrum was analysed with spectral models following the
Bayesian methodology of \citet{Buchner2014}. We only consider their
best model, \texttt{torus+pexmon+scattering}, which consists of 
\begin{enumerate}
\item an intrinsic power law spectrum modified by photo-electric absorption
and Compton scattering from an obscurer with toroidal geometry simulated
by the \texttt{torus} model of \citet{Brightman2011a}, 
\item a Compton-reflection component approximated by the \texttt{pexmon}
model \citep{Nandra2007} and 
\item a soft scattering component, which is parameterised by an unabsorbed
power law with the same spectral slope as the intrinsic one and normalisation
that cannot exceed 10\% of the intrinsic power-law spectrum. 
\end{enumerate}
This model has five free parameters, the slope of the intrinsic power-law
spectrum, $\Gamma$, the line-of-sight column density of the obscurer,
$N_{H}$, the normalisations of the intrinsic power-law, Compton-reflection
and soft scattering components. For the \texttt{torus} model of \citet{Brightman2011a}
we fix the opening angle parameter to $45\text{\textdegree}$ and
the viewing angle parameter to the maximum allowed, $87\text{\textdegree}$,
i.e. edge on. For \texttt{pexmon} component we use the same incident
power-law photon index as the intrinsic one spectrum and fix the inclination
to $60\text{\textdegree}$. The normalisation of the \texttt{pexmon}
($R$ parameter) is modelled relative to that of the \texttt{torus}
component and is allowed to vary between 0.1 and 100. In the analysis,
we use flat priors on the luminosity/normalisation parameters and
the column density, and an informed Gaussian prior of mean $1.95$
and standard deviation $0.15$ for the photon index $\Gamma$ \citep{Nandra1994}.

In the spectral analysis, redshifts are either fixed, when spectroscopy
is available, or are included in the form of priors that follow a
probability distribution function (photometric redshifts and missing
counter parts, see Section \ref{sec:redshifts} above).

Following the methodology of \citet{Buchner2014}, we use the $0.5-8\text{\,\ keV}$
spectrum and the redshift information to infer posterior distributions
for the X-ray luminosity $L_{X}$, redshift $z$ and column density
$N_{H}$.

\subsection{\label{sec:sample}Sample selection}

Based on the initial hard-band detection, normal galaxies (i.e. non-AGN)
should already be largely absent from our sample, except for strongly
star-bursting galaxies. To exclude these in a conservative manner,
we remove intrinsically faint sources that have $L_{X}<10^{42}\,\text{ erg/s}$
with 90\% probability. The number of objects used in the luminosity
function analysis is shown in Table \ref{tab:Steps}. Figure \ref{fig:Luminosity-redshift}
shows how the sample is spread in the luminosity-redshift plane.

\section{Methodology\label{sec:Methodology}}

In this work, we want to study the evolution of AGN obscuration. More
specifically, we are interested in the distribution of the population
in X-ray luminosity $L$ and column density $N_{H}$, i.e. the luminosity
function. Our sample is a specific draw from this distribution, with
a sampling bias in detection and incomplete information on each object.
Inference methods for such scenarios have been known for a long time
\citep{Marshall1983}.

\subsection{\label{sub:LF-intro-1}Luminosity function analysis\label{sub:LF-models}}

Our general approach to establish the evolution of the obscured and
unobscured populations of AGN is to determine the X-ray luminosity
function, accounting for all measurement uncertainties, which can
then be analysed as a function of obscuration to determine the absolute
and relative number densities of obscured and unobscured sources. 

Direct visualisation of the data is difficult if we want to stay true
to the uncertainties. Faint objects, which dominate the sample, are
highly uncertain in their properties, namely the intrinsic luminosity
$L$, obscuration $N_{H}$, but also their redshift $z$ in the case
of photometric estimates. For instance, Compton-thick AGN can have
large ``probability clouds'' for their parameters. This prohibits
us from assigning objects to bins for visualisation. For direct visualisation,
three approaches can be considered: (1) assigning each object to a
random luminosity bin based on its probability distribution, and then
estimating the density in the bin, (2) assigning each object to each
luminosity bin with a probability weight, and then estimating the
density in the bin, (3) computing for each luminosity bin the number
of objects that have a higher luminosity with e.g. 90\% probability.
Method (2) has the difficulty that the ``number'' in each bin is
no longer integer -- requiring interpolation of the Poisson distribution
formula. The methods (1) and (2) assume a frequency interpretation
of the uncertainty probability distributions -- which is not reasonable
as every object is different and the sample size is small. Method
(3) may be useful for checking whether data and model agree, but does
not yield an intuitive visualisation. We thus find none of these methods
satisfying, and develop a new approach (next section), which visualises
the data and estimates the luminosity function at the same time.

In the Appendix A, we review the statistical footing of analysing
population demographics by reviewing and combining the works of \citet{2004AIPC..735..195L}
and \citet{Kelly2008}. In this work, we use the usual Poisson likelihood:

\begin{eqnarray}
{\cal L} & = & \frac{\prod_{k}\int\frac{\phi({\cal C})}{d{\cal C}}\cdot p(d_{k},\, D|{\cal C})\cdot\frac{dV}{dz}\, d{\cal C}}{\exp\left\{ \int\frac{\phi({\cal C})}{d{\cal C}}\cdot A({\cal C})\cdot\frac{dV}{dz}\, d{\cal C}\right\} }\label{eq:likelihood}
\end{eqnarray}
where ${\cal C}=\{\log L,\, z,\,\log\, N_{H}\}$, $p(d_{k},\, D|{\cal C})$
represents the results of the spectral analysis of data $d_{k}$ from
the detected object $k$, which is weighted by the luminosity function
$\phi$. The integral in the denominator computes the expected number
of sources by convolving the luminosity function with the sensitivity
curve $A$ (shown in Figure \ref{fig:Area-curve} as a dotted magenta
line). The conversion from $L,\, z,\, N_{H}$ to sensitive area requires
a spectral model, for which we use the torus model, and we average
over $\Gamma$ in exact correspondence to the Gamma prior used in
our analysis. Here, the additional scattering components (\texttt{+pexmon+scattering},
see Section \ref{sec:spectra}) are not used. Both contribute only
minimally to the $2-10\text{ keV}$ flux: \citealp{Buchner2014} presents
an extreme example in detail, where both components show the strongest
contribution in their sample, but the $2-10\text{ keV}$ flux is only
affected by $10\%$ (0.04 dex). Future studies may take these components
and their normalisations into account as well. The Appendix A gives
an extensive derivation and caveats for equation \ref{eq:likelihood}.

\subsection{Non-parametric approach\label{sub:Non-parametric-approach}}

We would like to use the power and safety of a likelihood-based analysis
but without the rigidity of a functional form, allowing discovery
of the shape of the luminosity function. Our method for analysing
the luminosity function is thus, in a simplistic description, to fit
a three-dimensional ($L,\, z,\, N_{H}$) histogram as the luminosity
function model.

\begin{figure}[h]
\begin{centering}
\includegraphics[width=0.49\textwidth]{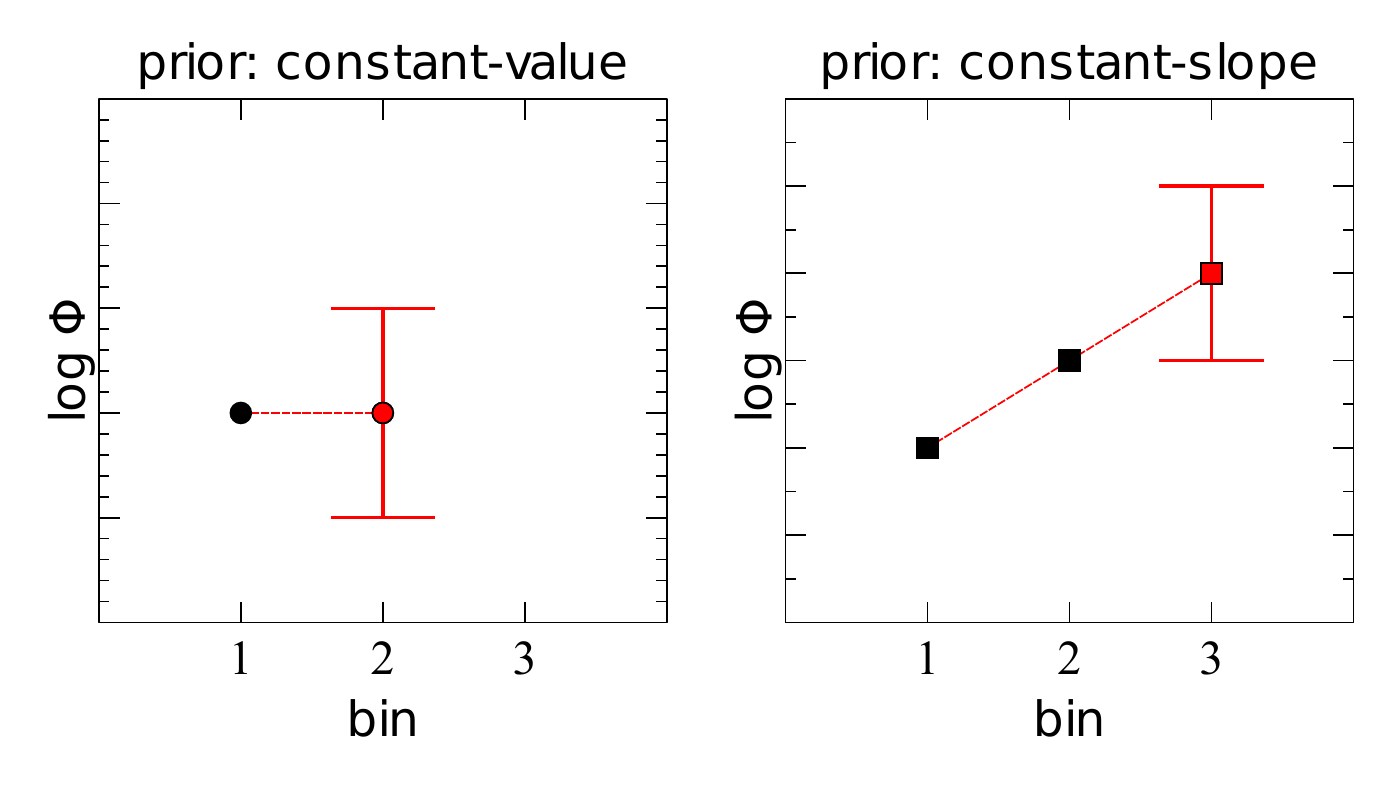}
\par\end{centering}

\protect\caption{\label{fig:priors}Illustration of the two smoothness priors used.
The \emph{left} \emph{panel} shows the ``constant-value'' prior:
The extrapolation from well-constrained step-function bins (black)
to a neighbouring bin (red) is done by assuming the same value with
a fixed uncertainty $\sigma$, whose value encodes the assumed correlation
strength or smoothness. The \emph{right panel} illustrates the ``constant-slope''
prior. The value for the neighbouring bin (red) is predicted by continuing
the same slope from the black points. As the prior is defined in logarithmic
units of the density, this behaviour corresponds to preferring a powerlaw.}
\end{figure}

Using e.g. $10\times10\times10=1000$ bins already means that the
problem is largely under-determined. We thus need to input additional
knowledge. Here we make the reasonable assumption that the function
does not vary rapidly between neighbouring bins (in particular for
the redshift). We encode this smoothness prior in two approaches,
both using the Normal distribution to penalise large deviations. These
two priors, \emph{``constant-value'' }and\emph{ ``constant-slope''}
are explained below and illustrated in Figure \ref{fig:priors}.
\begin{itemize}
\item \emph{``constant-value''}: This prior retains the current value
unless constraints are imposed by the data. The value of a bin should
scatter around its neighbour density value $\log\phi_{i+1}=\text{Normal}(\log\phi_{i},\,\sigma)$
with an allowed correlation width $\sigma$ for each axis ($\sigma_{L}$,
$\sigma_{z}$).
\item \emph{``constant-slope''}: This prior keeps power law slopes intact
unless constraints are imposed by the data. The log-density slope
between bins should scatter around its neighbours slope $\log\phi_{i+2}=\log\phi_{i+1}+\text{Normal}(\log\phi_{i+1}-\log\phi_{i},\,\sigma)$,
with the deviation from the slope for each axis ($\sigma_{L}$, $\sigma_{z}$). 
\end{itemize}

\begin{figure*}[t]
\begin{centering}
\includegraphics[width=1\textwidth]{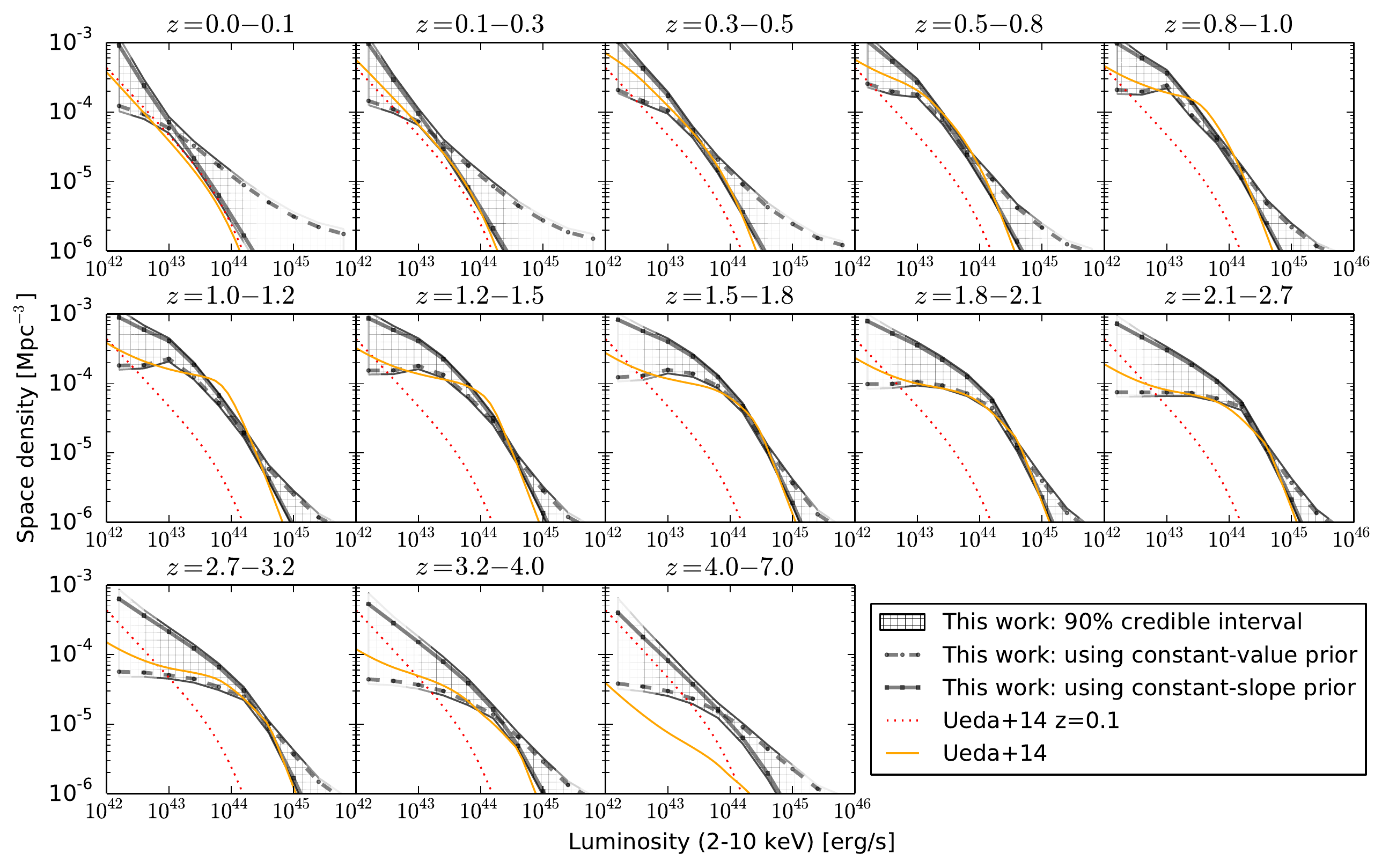}
\par\end{centering}

\protect\caption{\label{fig:totallf_L}Total X-ray luminosity function in the $2-10\text{\,\ keV}$
spectral band. Each panel corresponds to different redshift intervals.
We show our field reconstruction based on a step function in black.
The dashed black lines refer to the prior preferring to keep the space
density constant, while the solid black line prefers to keep the slope
of the density (in z and L) constant. Whenever the data constrain
the result well, the space density estimate from either prior prescription
(constant-value or constant-slope) is the same. Only when they extrapolate
away from any constraints, the results differ. We thus take the difference
in the reconstructions as an indication of whether the data or the
priors dominate the result. The hatched regions indicate our measure
of the uncertainty, using the 10\%-90\% quantiles of the posterior
samples from both priors together. The orange thin solid line shows
the reconstruction by \citet{Ueda2014}. The dotted red curve is their
local ($z=0.1$) luminosity function kept constant across all panels
for comparison. There are important differences between our luminosity
function estimates and that of \citealp{Ueda2014}, discussed in the
text.}
\end{figure*}

For the $N_{H}$ dimension, we always use the constant-value prior,
as in our view a power law dependence is not appropriate here. With
this simple prescription, we can recover a smooth field by fitting
a model, whose shape is driven by the data.

Whenever the data constrain the result well, the results from either
prior prescription (constant-value and constant-slope) will be the
same. Where the constraints are poor, the results will differ depending
on the adopted prior. We thus take the difference in the reconstructions
as a\change{}{n} indication of whether the data or the priors dominate,
and in the latter case the difference between the two is an indication
of the uncertainty in the determination.

It should be stressed that the choice of binning and in particular
the correlation strengths $\sigma_{L}$, $\sigma_{z}$, $\sigma_{N_{H}}$
can influence the result. Motivated by the number of data points in
each bin, we choose our pixelation as 11 bins of logarithmically spaced
luminosity $42-46$ (units erg/s), redshift bin edges 0.001, 0.1,
0.3, 0.5, 0.75, 1, 1.25, 1.5, 1.8, 2.1, 2.7, 3.2, 4, 7 and $\log\, N_{H}$
bin edges 20, 21, 22, 23, 23.5, 24, 26. The correlation strengths
$\sigma$ is defined for neighbouring bins, and their choice is important.
Notably, if $\sigma$ is too small, the model will be flattened out,
as the prior dominates, while if $\sigma$ is too large, no smoothness
assumption is used, and uncertainties will be large. Ideally, we would
like to recover $\sigma$ from the data themselves. Unfortunately,
our tests show that this computation is not numerically stable. We
do find however that above some value of $\sigma$, the results are
stable regardless of the choice of $\sigma$. We thus just choose
reasonable values for $\sigma$, namely $\sigma_{L}=0.5$, $\sigma_{z}=0.5$
and $\sigma_{N_{H}}=0.75$. This encodes, roughly speaking, that neighbouring
bins have the same order of magnitude in space density. These values
have been chosen after a few initial tests, but were not tuned to
give optimal results. Rather, we believe, they are one possible, and
reasonable, a priori choice.

We are also highly interested in the uncertainties of our smooth field
reconstruction method. Bins with data will be tightly constrained,
while bins without information will have increasing uncertainty with
distance from the data. Making uncertainty estimates with so many
parameters is not trivial, especially as the parameters are correlated
by definition. We use a Hamiltonian Markov Chain Monte Carlo code
named ``Stan'' (\citealp{stan-software:2014}). Stan uses the sophisticated
No-U-Turn Sampler (NUTS, \citealp{Hoffman2011NUTS}) technique to
ensure rapid mixing of the Markov chain by avoiding cyclic explorations.
Our Stan model is shown in Algorithm \ref{alg:Stan-code} in the Appendix.

\section{\label{sec:Results}Results}

\begin{table*}[t]
\begin{centering}
\protect\caption[Key statistics on the fraction of obscured AGN. ]{\label{tab:Key-statistics}Key statistics on the fraction of obscured
and Compton-thick AGN. }

\par\end{centering}

\centering
\begin{threeparttable}

\begin{centering}
\begin{tabular*}{0.9\textwidth}{@{\extracolsep{\fill}}rcccc}
\toprule[1.5pt] & Cosmic time average\tnote{a} & At $z=0$\tnote{a,b} & Maximum\tnote{a,c} & $z_{\text{Maximum}}$\tnote{c}\tabularnewline
\midrule
Obscured fraction ($>10^{22}\text{ cm}^{-2}$): & $\valuetotalobscspacedensitylowzfrac$ & $\valuetotalobscspacedensityaveragefrac$ & $\valuetotalobscspacedensitypeakzfrac$ & $>2.25$\tabularnewline
Compton-thick fraction ($>10^{24}\text{ cm}^{-2}$): & $\valueCTspacedensityaveragefrac$ & $\valueCTspacedensitylowzfrac$ & $\valueCTspacedensitypeakzfrac$ & (unconstrained)\\
\bottomrule[1.5pt] 
\end{tabular*}
\par\end{centering}

\begin{tablenotes}

\item[a] These fractions relative to the total space density of the
population are estimated by integrating the X-ray luminosity function
over cosmic time and within X-ray luminosity range $L(2-10\text{ keV})=10^{43.2}-10^{46}\text{ erg/s}$.
The uncertainties are computed using the \change{$5\%$ and $95\%$}{$10\%$
and $90\%$} quantiles of the posterior distribution, i.e. the true
value is bracketed with $90\%$ probability. 

\item[b] For this estimate we used our lowest redshift bin.

\item[c] The peak is computed by identifying the maximum value (fraction)
and location (redshift) in each posterior realisation, and considering
the distribution of each series. This leads only an upper limit for
the peak location for obscured AGN. For Compton-thick AGN, no upper
or lower limit could be determined. 

\end{tablenotes}

\end{threeparttable}
\end{table*}

A few words on the visualisation are warranted. In all relevant figures
we plot the posterior distributions of our three-dimensional step
function model from various axis views ($L$, $z$ and $N_{H}$).
We always show the median result for the constant-value prior as a
\emph{dashed line}, and the median result for the constant-slope prior
as a \emph{solid line}. A feature of our methodology is that uncertainties
are realistic and reflect regions of parameter space where data are
sparse. Whenever the data constrain the result well, the results from
either prior prescription (constant-value or constant-slope) will
be the same. Only when constraints from the data are poor, the results
differ. We thus take the difference in the reconstructions as a indication
of whether the data or the priors dominate the result. We plot the
10\%-90\% quantile \emph{hatched regions} as a measure of the uncertainty,
by taking together the posterior samples from both priors.

\subsection{Total luminosity function}

Figure \ref{fig:totallf_L} shows our non-parametric total X-ray luminosity
function, integrated along the $N_{H}$ axis. Large uncertainties
are visible at the bright end at low redshift ($z<0.3$), where our
sample is small due to the limited cosmological volume. At high redshift
($z>2$), the faint end becomes uncertain as low X-ray fluxes limit
the number of detected sources.

Our method captures the general shape, normalisation and evolution
with redshift of the X-ray luminosity function (XLF) as inferred or
assumed in previous parametric determinations. In Figure \ref{fig:totallf_L}
we show in particular the comparison with a recent comprehensive study
of the XLF by \citet{Ueda2014}. The overall shape of the luminosity
function is a double power-law with a break or bend at a characteristic
luminosity ($L_{*}$), the value of which increases with redshift.
As found in previous studies, the space density shows a rapid evolution
up to around $z\sim1$ at all luminosities, being most prominent at
high luminosities due to the positive evolution of $L_{*}$. We find
that the positive evolution continues up to $z\sim3$ \citep[as in][]{Aird2010}
above which the population starts to decline. It is important to emphasise
that in our non-parametric approach these features are imposed by
the data and not by any assumptions about the functional form of the
X-ray luminosity function. 

While the general behaviour of the population is similar, Figure \ref{fig:totallf_L}
also shows important differences compared to some previous parametric
studies, specifically that of \citet{Ueda2014}. \change{}{At the
highest redshift bin ($z=4-7$), the space density drops sharply towards
high redshifts in their XLF reconstruction \citep[see also][]{Kalfountzou2014,Vito2014,Civano2011}.
However, our reconstruction remains, as suggested by the priors, at
space densities comparable to the previous redshift bin $z=3-4$.
In our data, there seems to be no evidence of a steep decline with
redshift. The difference may be due to the large uncertainties in
the redshift estimates used in this work.} At intermediate redshifts,
\change{their XLF}{the \citet{Ueda2014} XLF} shows a sharp flattening
below a luminosity of around $10^{44}\text{\,\ erg/s}$, after which
it steepens again at the lowest luminosities probed by the study\change{
$10^{44}\text{\,\ erg/s}$}{}. This behaviour is most apparent at
redshifts $z\sim1-2$ and is only easy to parameterise using the Luminosity
Dependent Density Evolution model (e.g. \citet{Ueda2003,Hasinger2005,Silverman2008}).
Our analysis does not support such a behaviour. This is most apparent
in the redshift range $z=0.8-2.1$, where our non-parametric field
analysis requires a significantly larger space density of AGN in the
critical luminosity range of $10^{43-44}\text{ erg/s}$. Furthermore
we see no strong evidence of a change in the faint-end slope in the
individual panels of Figure \ref{fig:totallf_L}. This may bring into
question the need for LDDE to explain the form and evolution of the
XLF. As our focus in this paper is on the demographics and evolution
of obscuration, rather the luminosity function per se, we defer further
discussion of this point to later work.

\subsection{Obscured and Compton-thick fractions\label{sub:Obscured-and-Compton-thick}}

We now report on the fraction of obscured AGN by comparing the space
density above $N_{H}>10^{22}\text{ cm}^{-2}$ to the total. Our non-parametric
approach allows us to explore this fraction in a model independent
way by integrating the luminosity function over luminosity and cosmic
time. We computed these fractions using only the luminosity range
$\ensuremath{L=10^{43.2-46}\text{ erg/s}}$, i.e. without the lowest
three luminosity bins. The choice of the luminosity range for the
presentation of the results is to minimise uncertainties associated
with the typically looser constraints our methodology yields at the
faint-end of the X-ray luminosity function. At virtually all redshifts
the AGN space density is better determined at luminosities $L>10^{43.2}\text{ erg/s}$.
In Table \ref{tab:Key-statistics}, we find that the fraction of obscured
objects in the Universe is $75\%$, with narrow uncertainties. The
fraction of Compton-thick AGN ($N_{H}>10^{24}\text{ cm}^{-2}$) is
approximately $35\%$.

\begin{figure*}[t]
\begin{centering}
\includegraphics[width=0.49\textwidth]{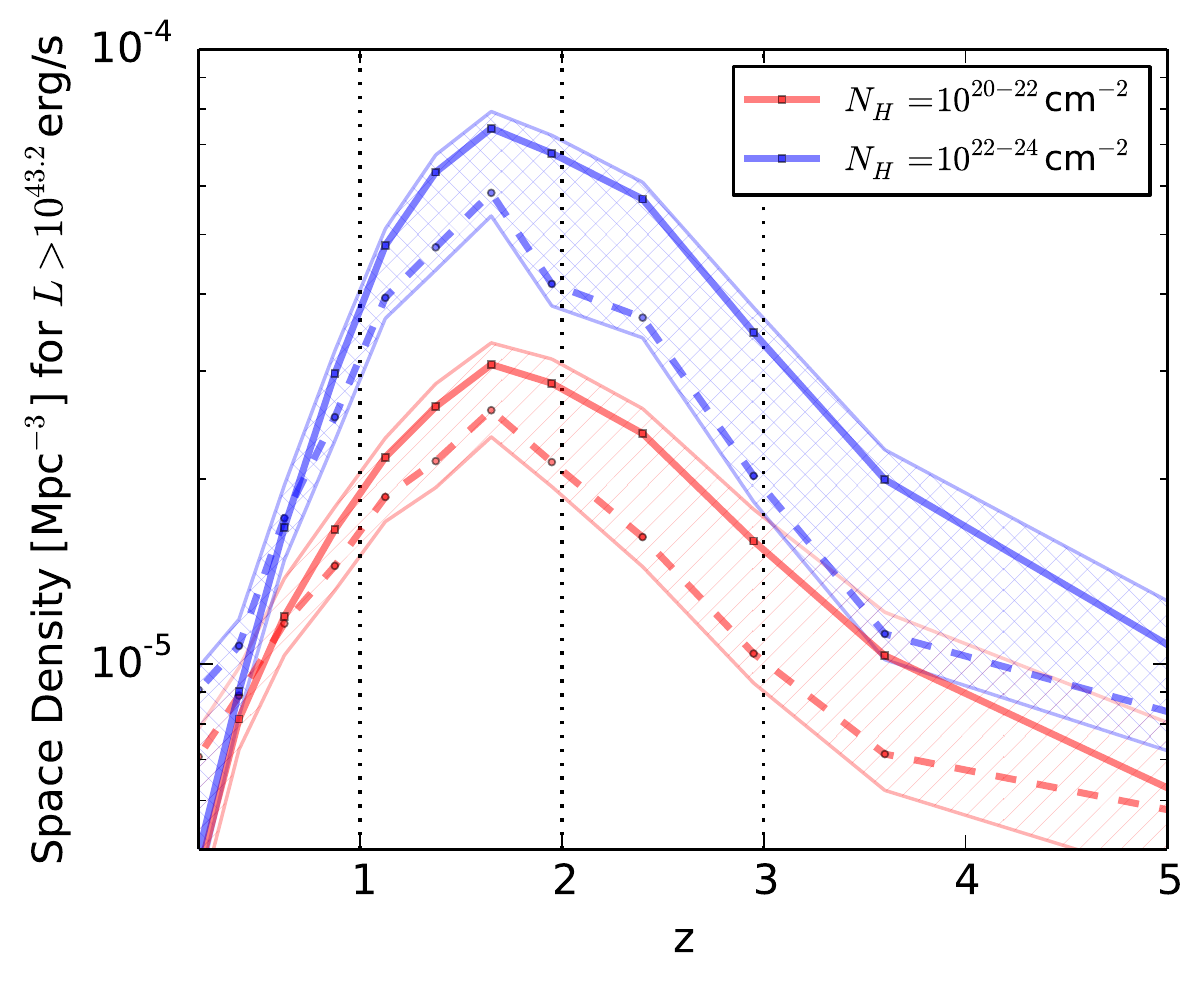}
\includegraphics[width=0.49\textwidth]{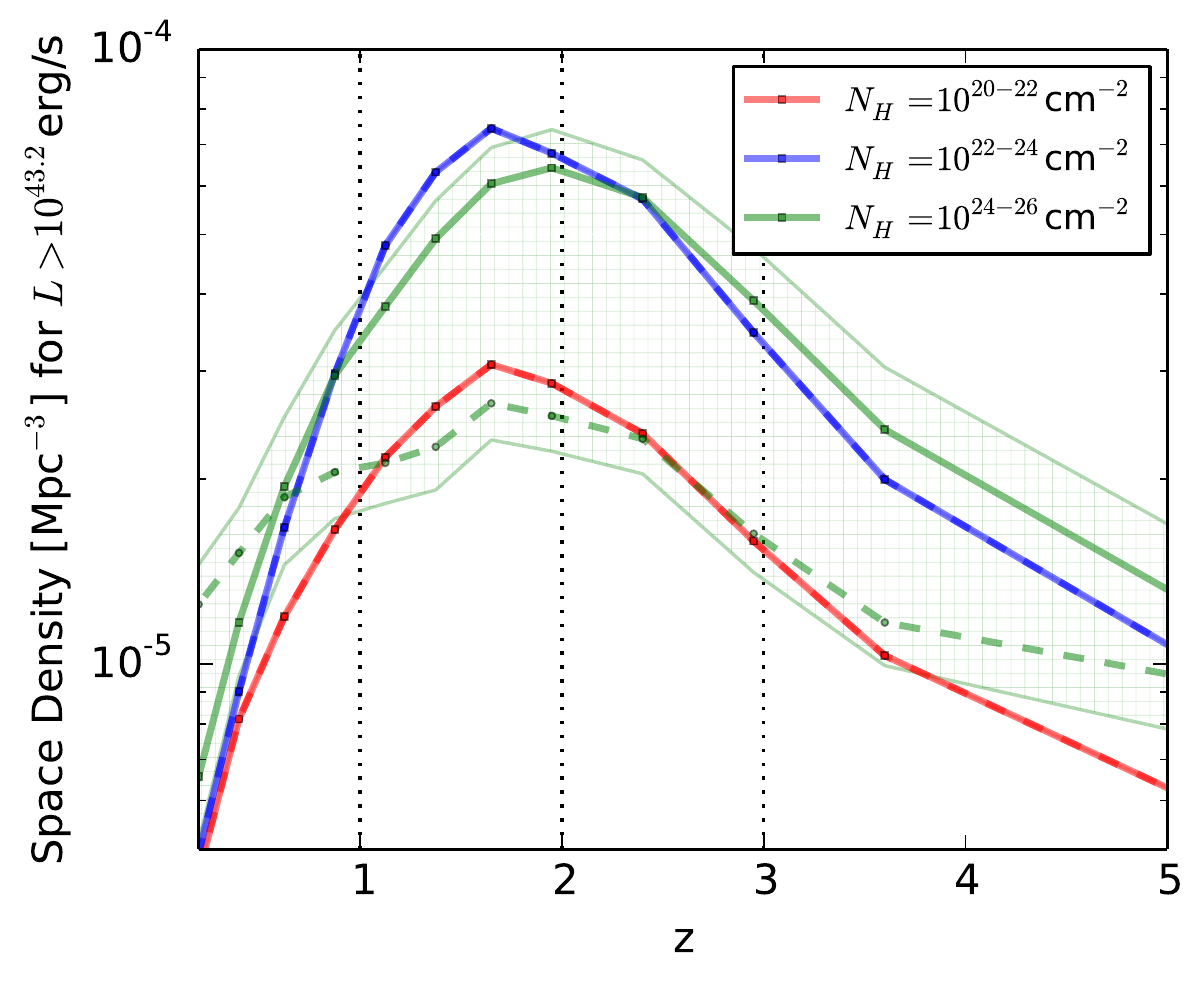}
\par\end{centering}

\protect\caption{\label{fig:fieldsummary_z}Redshift evolution of the space density
of AGN with $L_{X}>10^{43.2}\text{ erg/s}$, for various column densities.
Our step function reconstruction is represented by points at the bin
centre, which are connected by lines (dashed for the constant-value
prior, solid lines for the constant-slope prior). In the \emph{left
panel}, distinct evolutions for the Compton-thin obscured (blue shaded
region, top) and unobscured (red shaded region, bottom) can be observed.
The \emph{right panel} plots the evolution of Compton-thick AGN as
a green shaded region. To facilitate the comparison we also plot the
evolution of the unobscured and obscured Compton-thin AGN reconstruction
in the case of the constant-slope prior (solid lines). All AGN sub-populations
split by the level of obscuration experience similar space density
evolution, which can be described by a rise from $z=0.5$ to $z=1.25$,
a broad plateau at $z=1.25-2.1$ and a decline at higher redshift.
There is also evidence that moderately obscured, Compton-thin AGN
($N_{H}=10^{22}-10^{24}\,\text{ cm}^{-2}$) are evolving faster in
the redshift interval $0.5-4$ in the sense that they reach peak space
densities higher than the other AGN sub-populations. The space density
of Compton-thick AGN has the highest uncertainty, due to poor statistics
in the low-luminosity range ($L<10^{44}\text{ erg/s}$). Nevertheless,
there is tentative evidence that the evolution of Compton-thick AGN
is weaker than that of the Compton-thin obscured AGN (blue), and in
fact closer to the evolution of the unobscured AGN (bottom red solid
line).}
\end{figure*}

\begin{figure*}[t]
\begin{centering}
\includegraphics[width=1\textwidth]{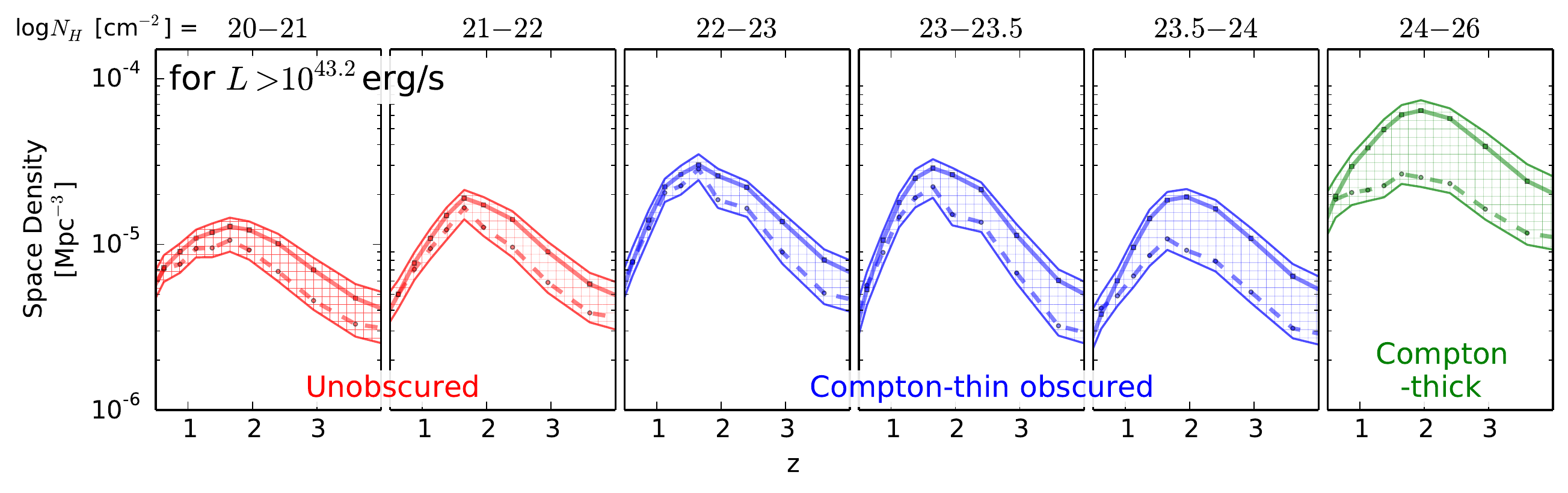}
\par\end{centering}

\protect\caption{\label{fig:Evolution-NH}Redshift evolution of space density of AGN
split by the level of obscuration. Different panels correspond to
different hydrogen column density interval as indicated at the top.
The plot highlights that the redshift evolution of the AGN space density
is strongest for the column density bins $N_{H}=10^{22}-10^{23}\text{ cm}^{-2}$
and $N_{H}=10^{23}-10^{23.5}\text{ cm}^{-2}$.}
\end{figure*}

Additionally, we test whether these fractions are constant over cosmic
time. This is done by noting the peak fraction and its redshift in
each posterior realisation of our reconstruction, and averaging the
results. \change{In the lower sections of Table \ref{tab:Key-statistics},
we report that at $z>2$, the obscured fraction tends to a higher
value of almost $85\%$.}{As indicated in Table \ref{tab:Key-statistics},
the obscured fraction tends to a higher value of almost $85\%$ at
$z>2$.} This shows that the fraction of obscured AGN varies through
cosmic time. In the following section, we investigate this evolution
of the obscured fraction further. For the Compton-thick fraction,
no peak can be identified as the redshift is unconstrained. This indicates
that the Compton-thick fraction is constant over cosmic time at approximately
$35\%$.

\subsection{Obscuration-dependent evolution\label{sub:Obscuration-dependent-evolution}}

To explore the evolution of the obscured AGN fraction further, Figure
\ref{fig:fieldsummary_z} plots the evolution of the space density
of unobscured ($N_{H}=10^{20-22}\,\text{ cm}^{-2}$), moderately obscured
($N_{H}=10^{22-24}\,\text{ cm}^{-2}$) and Compton-thick ($N_{H}=10^{24-26}\,\text{ cm}^{-2}$)
AGN with luminosities $L>10^{43.2}\text{ erg/s}$. In the left panel,
we find that moderately obscured and unobscured AGN follow similar
evolutionary patterns, namely an increase from $z=0$ to $z=1.2$
where their space density peaks and a decline at higher redshifts.
However, there are also differences. Moderately obscured AGN evolve
much faster from $z\sim0.5$ to $z\sim1.5$. 

The right panel of Figure \ref{fig:fieldsummary_z} shows the space
density of Compton-thick AGN. The evolution of the Compton-thick AGN
population has larger uncertainties than unobscured and moderately
obscured Compton-thin AGN. Nevertheless we find that their space density
shows a broad plateau at $z\approx1-3$, followed by a decline to
both lower and higher redshifts. The evolution of Compton-thick AGN
also appears weaker than that of moderately obscured Compton-thin
ones. This behaviour is contrary to the $N_{H}$ smoothness prior,
which prefers neighbouring $N_{H}$ bins to have the same value. Thus
it can be concluded that the data drive the result of different evolutions
for different obscurations: A strong obscured Compton-thin evolution
compared to a weaker evolution in both Compton-thick and unobscured
AGN.

The different evolution of AGN with different levels of obscuration
is further demonstrated in Figure \ref{fig:Evolution-NH}, where the
AGN sample is split into finer $N_{H}$ bins. All sub-populations
experience the same evolutionary pattern, a rise from redshift $z\approx0.5$,
a peak at $z\approx1.5$ and a decline at higher redshift. The AGN
that undergo the strongest evolution are those with columns densities
around \change{$N_{H}\thickapprox10^{23}\text{ cm}^{-2}$}{$N_{H}=10^{22-23.5}\text{ cm}^{-2}$}.
Both unobscured and Compton-thick AGN evolve less strongly.

\subsection{Luminosity-dependence and evolution of the obscured fraction\label{sub:lumdep-obsc}}

\begin{figure*}[t]
\begin{centering}
\includegraphics[width=1\textwidth]{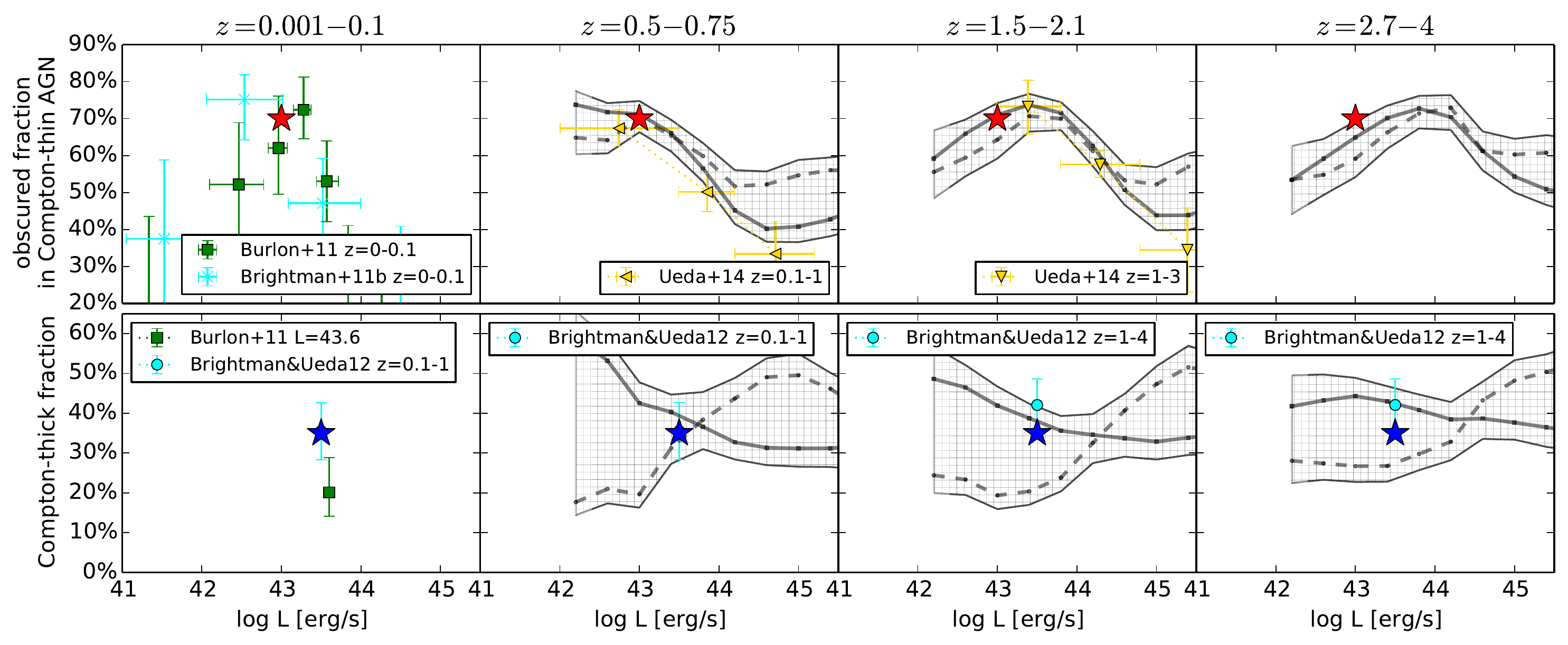}
\par\end{centering}

\protect\caption{Luminosity-dependence of the obscurer. The \emph{top row} plots \label{fig:obscured-fraction-lum}the
obscured fraction of Compton-thin AGN (CTNOF, eq. \ref{eq:CThin})
for various redshift intervals. The shaded grey region are the constraints
from our non-parametric method. Additionally, we compare to the work
of \citet{Ueda2014} (yellow points). For reference, the red star
symbol is placed at $70\%$ and $L=10^{43}\text{ erg/s}$ across the
panels. In the \emph{top left panel}, the results from local surveys
\citep{Burlon2011,Brightman2011b} which report a similar shape. The
CTNOF shows a distinct peak, which is placed at the red star for local
surveys, but appears to move to sequentially higher luminosities at
higher redshift. In the \emph{bottom row,} \label{fig:CT-fraction-lum}the
luminosity-dependence of the Compton-thick fraction (eq. \ref{eq:CT})
is plotted. Our results (shaded grey) show that the Compton-thick
fraction is compatible with being constant at $\sim35\%$ (blue star
symbol for reference at $L=10^{43.5}\text{ erg/s}$). For comparison,
previous surveys \citep{Burlon2011,Brightman2012} are shown (see
text).}
\end{figure*}

\begin{figure}[h]
\begin{centering}
\includegraphics[width=0.45\textwidth]{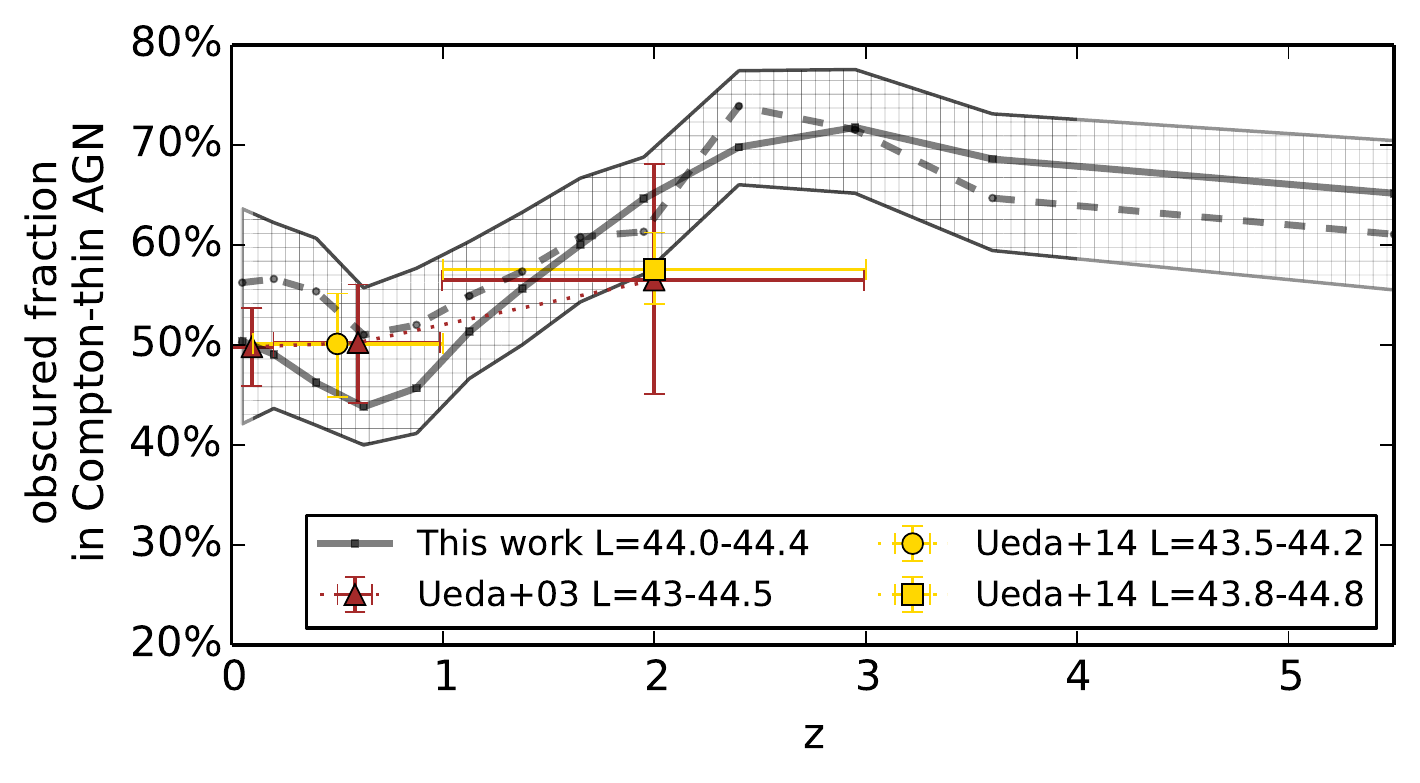}
\par\end{centering}

\begin{centering}
\includegraphics[width=0.45\textwidth]{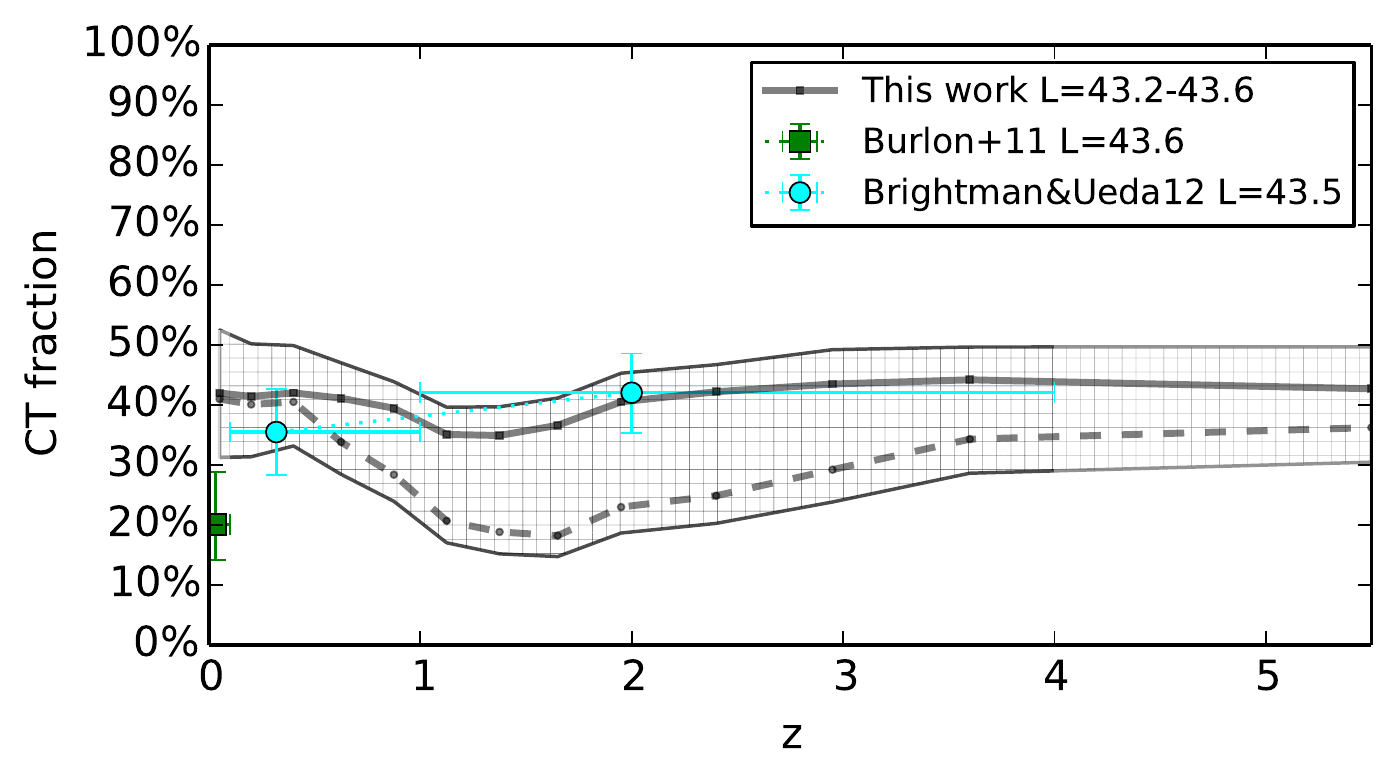}
\par\end{centering}

\protect\caption{Evolution of the obscured fraction.\emph{ Top panel}: \label{fig:obscured-fraction-evolution}The
evolution of the obscured fraction of Compton-thin AGN (CTNOF, eq.
\ref{eq:CThin}) is plotted as a grey shaded region for $L=10^{44}\text{ erg/s}$.
At $z\sim3$, the obscured fraction was higher ($75\%$) than today
($50\%$). For comparison also plotted are the constraints from \citet{Ueda2003}
(brown triangles) and \citet{Ueda2014} (yellow points).\protect \\
\emph{Bottom panel}: \label{fig:CT-fraction-evolution}The evolution
of the Compton-thick fraction at $L=43.5$. Our results are compatible
with a constant Compton-thick fraction of $\sim35\%$. We compare
our results to \citet{Brightman2012} (cyan circles). Local hard X-ray
Swift/BAT observations (green squares, \citealp{Burlon2011} for $z<0.1$)
derived a lower Compton-thick fraction of $\sim20\%$.}
\end{figure}

Previous studies suggest that the fraction of obscured AGN is a function
of accretion luminosity \citep[e.g.][]{Lawrence1991,Ueda2003,Simpson2005,Akylas2006,Silverman2008,Burlon2011,Ueda2014}.
The luminosity-dependence of the obscured fraction is thought to be
related to the nature of the obscurer, e.g. reducing it in physical
extent. In this section we investigate this issue using our model
independent approach. To illuminate the behaviour of the obscured
population we analyse the behaviour of Compton-thin and Compton-thick
AGN separately. This section only considers the obscured AGN that
are Compton-thin ($N_{H}<10^{24}\text{ cm}^{-2}$). To this end, we
define a new quantity, namely the obscured fraction of the Compton-thin
AGN (Compton-thin obscured fraction, CTNOF)

\begin{equation}
\text{CTNOF}:=\frac{\phi[N_{H}=10^{22-24}\text{ cm}^{-2}]}{\phi[N_{H}=10^{20-24}\text{ cm}^{-2}]}.\label{eq:CThin}
\end{equation}
In the top row of Figure \ref{fig:obscured-fraction-lum}, we find
that the CTNOF is a strong function of luminosity. There is evidence
for a peak at a certain luminosity and decline at both brighter and
fainter luminosities (the red star symbol provides a constant reference
point). Interestingly, the luminosity where the obscured AGN fraction
peaks appears to be a function of redshift. With increasing redshift,
the drop of the CTNOF at bright luminosities occurs at higher luminosities.
These results are in some agreement with the recent analysis in \citet{Ueda2014}.
One difference however is at the brightest luminosities, where our
non-parametric method tends towards a higher value ($50\%$). \change{}{Similar
high obscured fractions were suggested by \citet{Iwasawa2012a} and
\citet{Vito2014}, as opposed to 20\% in the local Universe \citet{Burlon2011}}.
\citet{Ueda2014} have better statistics at bright luminosities compared
to our work because they included more wide-are\change{}{a} and
shallow survey fields in the analysis. It is therefore likely that
our sample has poor statistics at the brightest luminosities which
makes the preference of the $N_{H}$ prior towards equipartition apparent.
The strength of our data is rather at the faint-end of the luminosity
function for moderate and high redshifts. There we find a significant
turnover at low luminosities (right top panel of Figure \ref{fig:obscured-fraction-lum}).
This result is independent of the adopted form of the prior, indicating
that this behaviour is strongly imposed by the data. A similar behaviour
of a peak luminosity and a turnover have been found in local samples
\citep{Burlon2011,Brightman2011b} (top left panel of Figure \ref{fig:obscured-fraction-lum},
corrected assuming a constant 20\% Compton-thick fraction in their
sample). The position of the peak they find is consistent with our
results at low redshift (red star symbol, $L=10^{43}\text{ erg/s}$).

When considering the obscured fraction at $L=10^{44}\text{ erg/s}$,
the evolution of the peak is imprinted as a rise with increasing redshift.
This is shown in Figure \ref{fig:obscured-fraction-evolution}, with
the same data points from literature. Observers considering mostly
AGN with luminosity $L\geq10^{44}\text{ erg/s}$ will see an increase
in the fraction of obscured AGN with redshift as shown here, while
observers considering intrinsically faint AGN ($L\leq10^{43}\text{ erg/s}$)
would observe the opposite trend (see Figure \ref{fig:obscured-fraction-lum}).

\subsection{Evolution of Compton-thick AGN\label{sub:results-CT}}

\begin{figure}[h]
\begin{centering}
\includegraphics[width=8.5cm]{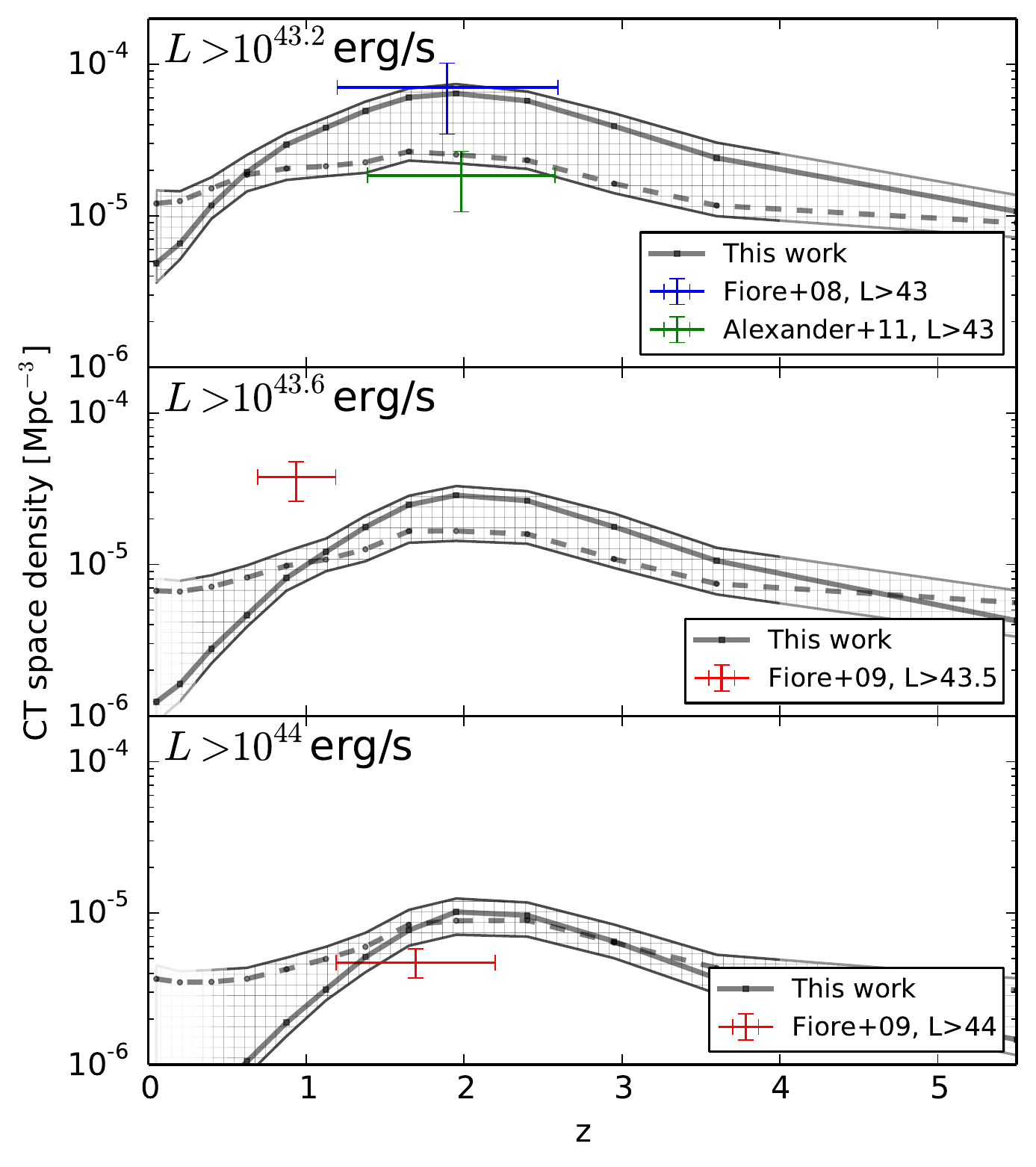}
\par\end{centering}

\protect\caption{\label{fig:CT-evolution}The evolution of the density of Compton-thick
AGN. Different panels correspond to X-ray luminosities brighter than
$10^{43.2}$ (\emph{top}), $10^{43.6}$ (\emph{middle}) and \change{$10^{43.4}$}{$10^{44}$}
(\emph{bottom}) in units of erg/s. Comparing our reconstruction (black)
to previous results, we find overall good agreement. The result of
\citet{Fiore2009} using the $L>10^{43.5}\text{ erg/s}$ cut is an
exception. Their estimate lies higher compared to our results, potentially
due to starburst galaxy or moderately obscured AGN contaminating their
sample (see text).}
\end{figure}

Compton-thick objects have been hypothesised to play a major role
in the accretion phase of AGN. The identification of such sources
in current X-ray surveys by XMM-Newton and Chandra is challenging
and therefore their space density as a function of redshift has remained
controversial. It is therefore important to place constraints on their
space density \change{}{of these sources} in the context of previous
studies. Figure \ref{fig:CT-evolution} shows the evolution of the
space density evolution of Compton-thick sources inferred by our non-parametric
methodology. Three luminosity cuts are shown which allow direct comparison
with previous studies \citep{Fiore2008,Fiore2009,Alexander2011},
which select Compton-thick AGN at infrared wavelengths. Figure \ref{fig:CT-evolution}
shows that the general trend is a decline of the space density of
this population with decreasing redshift. \change{This behaviour
appears to be reversed at redshifts above $z=2$.}{Above $z=2$,
a decline appears towards increasing redshifts.} Also, our results
are in rough agreement with previous estimates \citep{Fiore2008,Alexander2011}.
The results of \citet{Fiore2009} on moderate luminosity ($L\geq10^{43.5}\text{ erg/s}$)
Compton-thick AGN in the COSMOS field is an exception. The Compton-thick
AGN space density determined in that study is significantly higher
than our result. This may be attributed to contamination of obscured
AGN samples selected in the infrared by either dusty star-burst or
moderately obscured Seyferts \citep[e.g.][]{Georgakakis2010,Donley2010}.

To study the contribution of Compton-thick AGN to the total accretion
luminosity output it is necessary to characterise the luminosity-dependence
of Compton-thick AGN. The bottom panel of Figure \ref{fig:obscured-fraction-lum}
plots the Compton-thick fraction, defined as

\begin{equation}
\text{Compton-thick fraction}:=\frac{\phi[N_{H}=10^{24-26}\text{ cm}^{-2}]}{\phi[N_{H}=10^{20-26}\text{ cm}^{-2}]},\label{eq:CT}
\end{equation}
as a function of X-ray luminosity. Within the uncertainties we do
not find evidence for a luminosity dependence. If the constant-slope
prior is preferred (solid line), a similar behaviour as in the obscured
fraction is allowed. However, our results are consistent with a constant
Compton-thick fraction at $\sim35\%$. This is the first time the
luminosity-dependence of the Compton-thick fraction is constrained.

In Figure \ref{fig:obscured-fraction-evolution}, the bottom panel
shows the evolution of the Compton-thick fraction at $L=10^{43.5}\text{ erg/s}$.
Our results show a minor dip in the $z\sim1-2$ range. This is due
to the strong peak in Compton-thin sources (see above), causing a
strong rise in the denominator of the fraction. The blue data point
at $z<0.1$ was taken from the Swift/BAT analysis of \citet{Burlon2011},
where a considerably lower Compton-thick fraction was found, in disagreement
to our findings. This may be because of the small volume of our sample
at those redshifts, or differences in the analysis and in particular
the methods adopted to correct for incompleteness of the AGN samples
in the Compton-thick regime. Figure \ref{fig:obscured-fraction-evolution}
also compares our results with those of \citet{Brightman2012} in
the Chandra Deep Field South (cyan). Within the uncertainties we find
broad agreement. However, \citet{Brightman2012} argued in favour
of a redshift evolution (increase) of the Compton-thick fraction based
on their constraints and those of Swift/BAT \citep{Burlon2011}. In
our result we find no evidence for such a trend. Our uncertainties
are compatible with a constant Compton-thick fraction of $\sim35\%$.

\subsection{The contribution of Obscured AGN to the luminosity density\label{sub:results-CT-role-1}\label{sub:results-obsc-role}}

\begin{figure}[h]
\begin{centering}
\includegraphics[width=8.5cm]{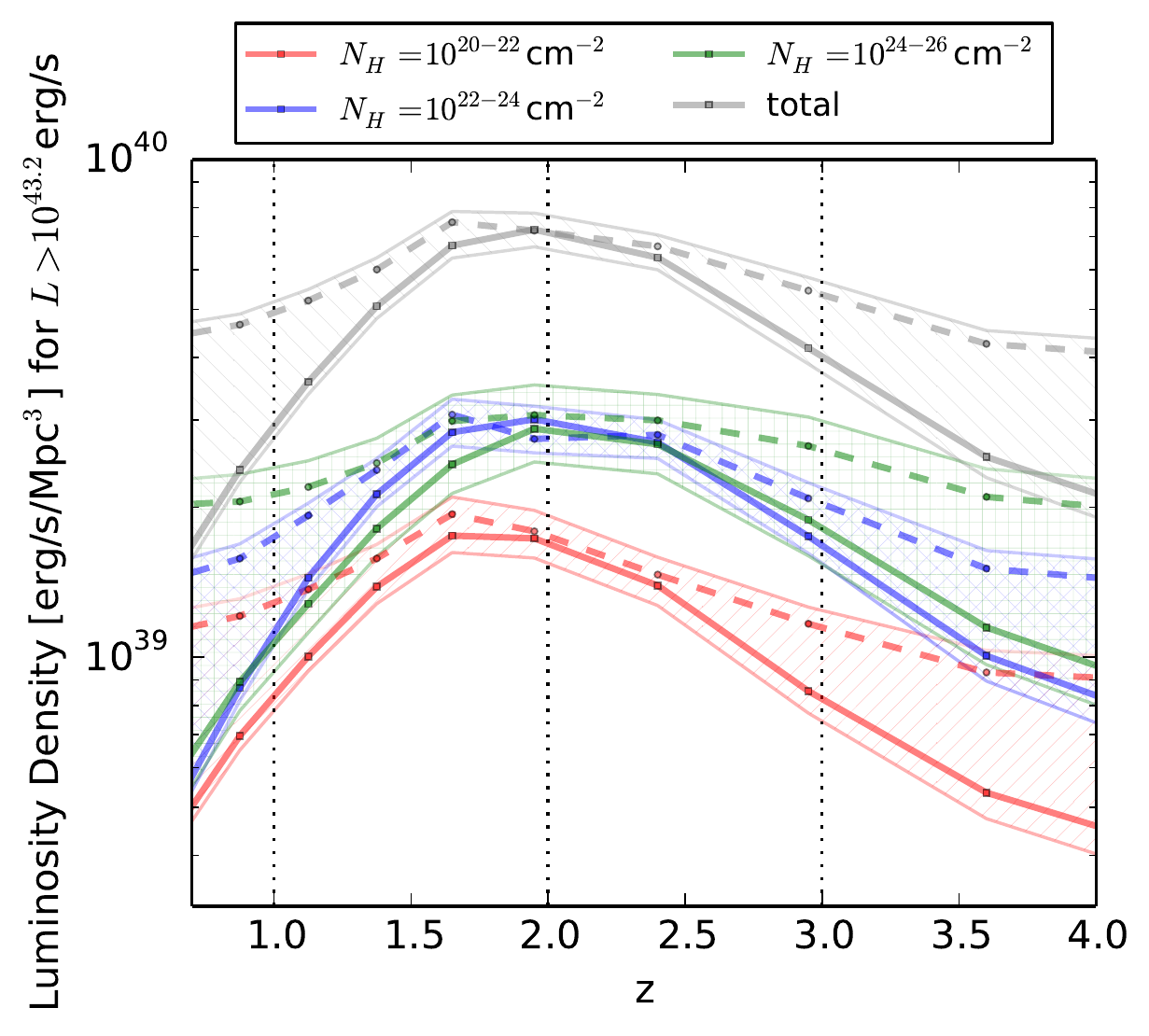}
\par\end{centering}

\protect\caption{\label{fig:fieldsummary_z_moment}Evolution of the X-ray luminosity
density of AGN with $L_{X}>10^{43.2}\text{ erg/s}$, for various column
densities. The luminosity output of AGN experiences a rise and fall
in density in the $z=1-3.5$ range (total as top gray shaded region).
The strongest contribution to the luminosity density is due to obscured,
Compton-thin (blue shaded region) and Compton-thick AGN (green shaded
region), which contribute in equal parts to the luminosity. In contrast,
the emission from unobscured AGN (red shaded region, bottom) is distinctively
smaller.}
\end{figure}

Due to the difficulty in capturing the entire population of AGN (see
previous section), the accretion history of SMBHs has remained uncertain
in the literature. Through the relation of mass accretion rate in
black holes to luminosity, the total luminosity output of AGN can
be used to study the total accretion history of the Universe. We compute
the total X-ray luminosity output at any particular redshift, i.e.
the luminosity density, by integrating the luminosity function as
$\int L_{X}\cdot\phi\, d\log L_{X}$. This is shown in the top of
Figure~\ref{fig:fieldsummary_z_moment} (grey). Here, we again use
$L>10^{43.2}\text{ erg/s}$, which captures essentially all of the
luminosity density and focus on the redshift range $z=0.5-4$ where
our data coverage is best. The luminosity density shows a broad peak
in the $z=1-3$ range, with a decline to both high and low redshifts. 

Additionally, in Figure~\ref{fig:fieldsummary_z_moment} we probe
the contribution of AGN split by their obscuration (Unobscured, Compton-thin
obscured, Compton-thick). One of our most striking results is that
Compton-thick AGN and obscured Compton-thin AGN contribute in equal
parts to the luminosity density in the $z=1-3$ range. These obscured
AGN contribute the majority ($\valuetotalobsclumdensityaveragefrac$),
while unobscured AGN only play a minor role. The shapes in evolution
are fairly similar. However, the evolution of unobscured AGN is weaker
compared to the Compton-thin obscured AGN when considering the $z\sim1-1.5$
range.

\subsection{$N_{H}$ distribution}

\begin{figure}[h]
\begin{centering}
\includegraphics[width=0.99\columnwidth]{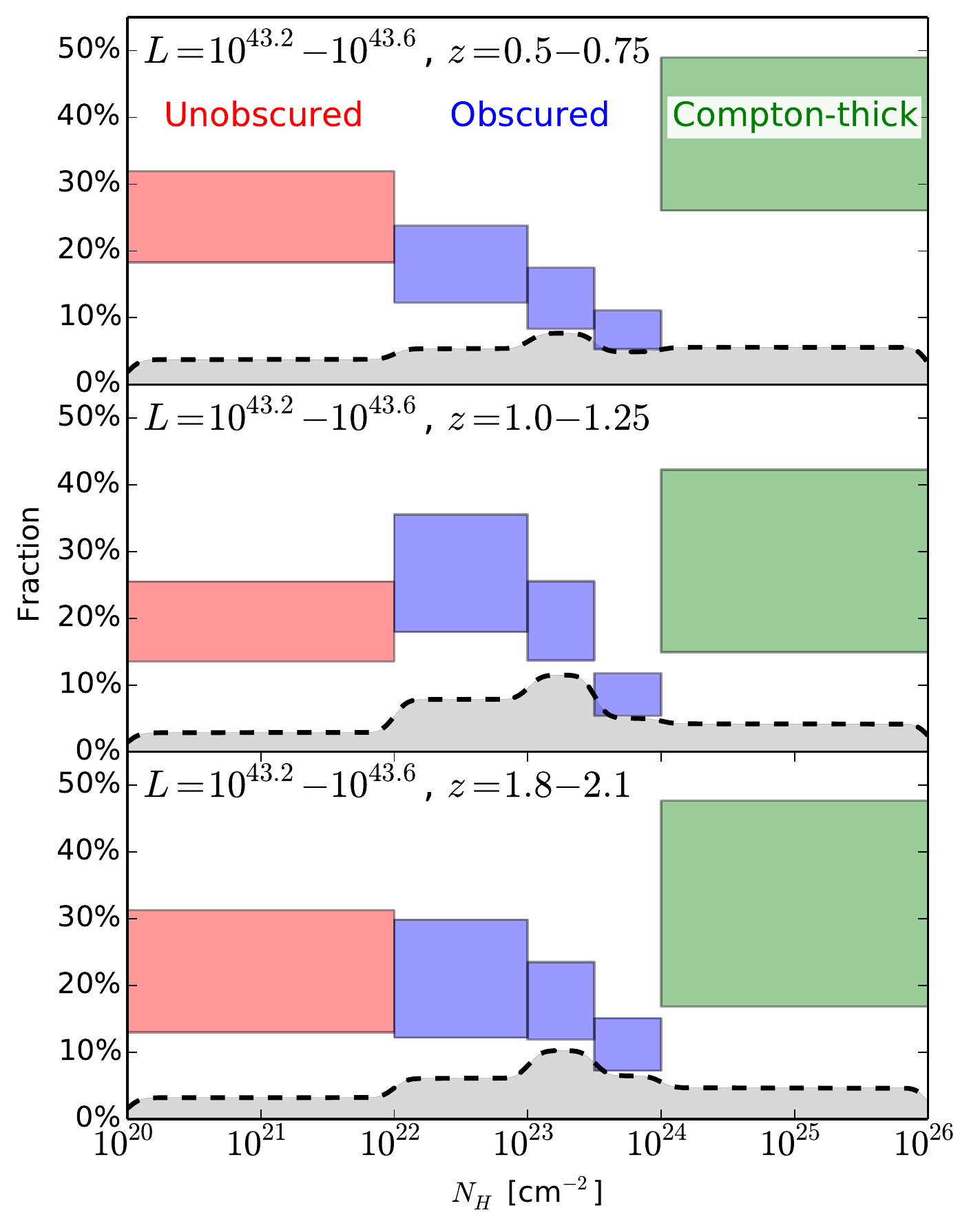}
\par\end{centering}

\protect\caption{\label{fig:NH-dists}Column density distributions at various redshifts.
Based on our field reconstruction, we compute the fraction of sources
in column density bins (red: ``unobscured'', blue: Compton-thin
obscured, green: Compton-thick). The boxes show the credible regions
containing 90\% of the posterior probability. \change{}{The dashed
line illustrates a possible $N_{H}$ density distribution that fits
these fractions.}}
\end{figure}

This medium luminosity interval, $L\approx10^{43.5}\text{ erg/s}$,
where most of the evolution occurs, is also where the obscuration
peaks according to Figure \ref{fig:obscured-fraction-lum}. To further
illustrate the evolution there, we look at how the column density
distribution evolves. Figure \ref{fig:NH-dists} plots the intrinsic
$N_{H}$ distribution for three redshift intervals in panels. Here,
the boxes indicate the upper and lower 90\% quantile on the fraction
of sources in the respective $N_{H}$ bin. The dashed line illustrates
a possible $N_{H}$ density distribution that fits these fractions.
When comparing the top panel to the middle panel in Figure \ref{fig:NH-dists}
in particular, the main effect appears to be that the fractions at
$N_{H}\approx10^{23}\text{ cm}^{-2}$ increase towards higher redshifts.

\section{\label{sec:Discussion}Discussion}

We combined deep and wide-area X-ray surveys conducted by Chandra
and XMM-Newton to initially constrain the space density of X-ray selected
AGN as a function of accretion luminosity, obscuring column density
and redshift. Our methodology accounts for the different sources of
errors. For example, we include in the analysis uncertainties, both
random and systematic, associated with photometric redshift measurements.
We also account for the lack of redshift information for sources without
optical or infrared counterparts. For the determination of the obscuration
and intrinsic luminosity of individual sources we use the X-ray spectral
analysis method of \citet{Buchner2014}, which takes into account
both the Poisson errors of X-ray spectra and photometric redshift
errors. The spectral analysis also uses a physically meaningful, multi-component
model that is consistent with recent ideas and observations on dominant
AGN emission processes and the structure of matter in the vicinity
of active supermassive black holes. Another important feature of the
analysis presented in this paper is the non-parametric method developed
to determine the X-ray luminosity function of AGN. This allows us
to explore how the space density of AGN depends on luminosity, redshift,
and column density without imposing any model. This frees us from
any assumptions on the dependence of the AGN space density to luminosity,
redshift, and column density.

We find that obscured AGN ($N_{H}>10^{22}\text{ cm}^{-2}$) dominate
the population of active super-massive black holes at all redshifts.
This is in agreement with previous investigations \citep[e.g.][]{Ueda2003,Akylas2006,Ueda2014,Merloni2011},
although our analysis highlights this point more robustly and quantitatively
(with uncertainties). We show that about $75\%$ of the AGN space
density, averaged over redshift, corresponds to sources with column
densities $N_{H}>10^{22}\text{ cm}^{-2}$ (see Table \ref{tab:Key-statistics}).
The contribution of obscured AGN to the accretion density of the Universe
is similarly large ($75\%$). The bulk of the black hole growth across
cosmic time is therefore taking place behind large column densities
of gas and dust clouds.

\subsection{The role of Compton-thick AGN\label{sub:discussion-CT-role}}

We are also able to place constraints on the number fraction of the
most heavily obscured, Compton-thick sources to the AGN population,
finding it to be $\valueCTspacedensityaveragefrac$ of the total population.
Results from previous AGN surveys have been divergent due to difficulty
of identifying Compton-thick AGN (e.g. $15-20\%$ in \citealp{Akylas2009},
$5-20\%$ in \citealp{Burlon2011} for the analyses of Swift/BAT surveys),
and relatively low compared to the requirements from the X-ray background
(see below). Our constraints on the Compton-thick fraction are in
good agreement with the estimates of \citet{Brightman2012} of $\sim35-40\%$.
However in their work they concluded an evolution of the Compton-thick
fraction by contrasting their high-redshift data to a local survey,
which reported a significantly lower fraction \citep[specifically][]{Burlon2011}.
In our analysis, we do not find any evidence for a redshift evolution
of the Compton-thick fraction. However, our sample lacks large, shallow
fields that can probe the local universe, and thus our results may
also benefit from being combined with a local estimate. For instance,
the $2-10\,\text{ keV}$ X-ray based work of \citet{Risaliti1999}
estimated that $50\%$ of all Seyfert 2 are observed with a Compton-thick
line of sight. If this estimation is taken, it is in agreement with
our results without the need for any evolution. However, hard X-ray
surveys have reported significantly lower values, e.g. $20_{-6}^{+9}\%$
in \citet{Burlon2011}\changed{, $9-17\%$ in \citep{Bassani2006,Malizia2012,Vasudevan2013a}}.
This shows that estimates for a local Compton-thick fraction have
been diverse, making it difficult to make any claim of a evolutionary
trend. \change{}{One possible source of uncertainty is the sensitivity
to sources with $N_{H}=10^{25-26}\text{ cm}^{-2}$. In this work we
have explicitly assumed that their space density is the same as those
of sources with $N_{H}=10^{24-25}\text{ cm}^{-2}$. The sparse sampling
(only 3 secure objects in our entire sample) of these heavily buried
population prohibits strong inferences.}

For discussing the accretion luminosity onto obscured AGN it is noteworthy
that the fraction of Compton-thick AGN does not show a luminosity-dependence
(Figure \ref{fig:CT-fraction-lum}). When considering currently accreting
AGN, Compton-thick AGN appear to play an invariant role with luminosity
and cosmic time. However, the large uncertainties in the luminosity-dependence
do not allow firm conclusions. As Figure \ref{fig:fieldsummary_z}
shows, the evolution of Compton-thick AGN may follow the unobscured
AGN closely in shape, rather than the Compton-thin obscured AGN which
evolve strongly. In the former case Compton-thick AGN can be modelled
as $35\%/25\%=1.4$ times more abundant than unobscured AGN over cosmic
time and luminosity. In the latter case, they would be approximately
equal in abundance and follow the luminosity dependence and its evolution
(see Section \ref{sub:lumdep-obsc}, discussed below in Section \ref{sub:discuss-CTNOF}).

Measuring the luminosity density allows a view on the importance of
obscuration with regard to the accretion history of SMBHs. \citet{Soltan1982}
originally argued that a major contribution to the accretion has to
be accumulated in obscured AGN, which were missing in surveys at the
time. Albeit obscured AGN have been detected in great numbers since
then, the same argument has been extended to Compton-thick AGN, whose
fraction has remained controversial. Our analysis shows the accretion
growth of SMBHs is indeed dominated by Compton-thick and Compton-thin
AGN ($\valuetotalobsclumdensityaveragefrac$), while unobscured AGN
only play a minor role. However, we also find that Compton-thick and
Compton-thin AGN each contributing half of the luminosity output (see
Figure \ref{fig:fieldsummary_z_moment}), with the Compton-thick AGN
contributing $\valueCTlumdensityaveragefrac$. This shows that Compton-thick
AGN are an important contributor to the accretion history. Our analysis
places, for the first time, tight constraints on the Compton-thick
contribution. Results on the Compton-thick fraction from studies of
the cosmic X-ray background have varied between 9\% \citep{Treister2009},
45\%-30\% \citep[luminosity-dependent]{Gilli2007} and 28\%-60\% \citep[not constrained]{Ueda2014},
while \citet{Shi2013} reported 38\%. This scatter is also due to
the fact that XRB fitting involves other parameters which are degenerate
with the Compton-thick fraction (e.g. additional reflection\change{}{,
see \citealp{Akylas2012}}). \change{}{It is worth emphasizing that
our estimate on the Compton-thick fraction is higher than previous
survey studies, but in agreement to those values assumed in X-ray
background synthesis analyses.}

The overall evolution of the accretion density for the AGN population
is presented in Figure \ref{fig:fieldsummary_z_moment}. There, the
peak is in a broad plateau at $z=1-3$, with a drop-off towards high
redshift. This trend may corroborate claims that the accretion onto
SMBHs correlates with the star-formation history of galaxies \citep[see][]{Merloni2011}.
Unlike previous works our results do not a-priori assume any form
of the evolution at high redshift.

Having constrained the Compton-thick fraction well, we can speculate
on the origin of this obscuration. The fraction of Compton-thick sources
is considerably large ($35\%$), necessitating that a large fraction
of viewing angles is obscured ($\sim20\text{\textdegree}$, see Figure
\ref{fig:Cartoon-obscfrac}). \change{Such }{}Compton-thick obscuration
is associated with the torus, as such column densities are not typically
reached by galactic gas. As a Compton-thick, smooth obscuration would
be unstable \citep{Krolik1988}, the current working hypothesis is
that the obscuration comes in discrete clouds. Under this clumpy torus
model obscured views are produced when clouds are encountered in the
line of sight. If the Compton-thin obscuration is also associated
with the torus, Compton-thick AGN are a simple extension of Compton-thin
obscured AGN. In principle it would also be possible then that Compton-thick
AGN have just a larger number of clouds in a clumpy torus. In this
case, their cumulative line-of-sight obscuration would provide a Compton-thick
view. Alternatively, these objects may have denser clouds overall.
Another possibility is that the cloud density is distributed such
that both regimes are covered. For further research in this regard,
the unbiased column density distributions shown in Figure \ref{fig:NH-dists}
can provide observational constraints for clumpy torus models. The
alternative scenario is that Compton-thin obscuration is associated
with a medium of different extent than Compton-thick obscuration (e.g.
through galactic and nuclear gas\change{}{, \citealp{Matt2000}}).
We discuss both kind of models in Section \ref{sub:discuss-CTNOF}
when also considering the luminosity dependence.

\subsection{Obscuration-dependent evolution\label{sub:discussion-obscured}}

In Figure \ref{fig:fieldsummary_z} we show that the space density
of obscured Compton-thin AGN experience a much stronger rise in the
redshift interval $z=0.5-1.2$ compared to unobscured AGN. This demonstrates
an observed obscuration-dependent evolution. Our method explicitly
contains a smoothness prior, preferring that the space density of
AGN as a function of column densities is the same. In contrast, the
results show strong differences in the evolution of unobscured and
obscured Compton-thin AGN. We can thus conclude that this result was
driven by the data.

In Figure \ref{fig:Evolution-NH} we find that the fastest evolving
population is the one with column densities $N_{H}=10^{22}-10^{23.5}\text{ cm}^{-2}$.
A possible interpretation of these trends is that different levels
of obscuration correspond to media with different spatial extents.
We investigated this further by focusing on the behaviour of moderately
obscured AGN ($N_{H}=10^{22}-10^{24}\text{ cm}^{-2}$).

\subsection{Luminosity-dependence and evolution of the obscured fraction\label{sub:discuss-CTNOF}}

An important result from our analysis is that the obscured Compton-thin
AGN fraction (CTNOF, eq. \ref{eq:CThin}) depends on both luminosity
and redshift (see Figure \ref{fig:obscured-fraction-lum}). The luminosity
dependence can be described by a peak of the CTNOF at a given luminosity
followed by a decline at both brighter and fainter luminosities. The
redshift dependence is manifested by a shift to brighter luminosities
of the peak of the obscured AGN fraction with increasing redshift.
The shift of this peak causes observers considering mostly AGN with
luminosity $L\geq10^{44}\text{ erg/s}$ to see an increase in the fraction
of obscured AGN with redshift (as shown in Figure \ref{fig:obscured-fraction-evolution}).
Previous studies also find that the fraction of obscured AGN decreases
with increasing luminosity \citep[e.g.][]{Ueda2003,Akylas2006,Ueda2014}.
When we consider a luminosity range where we have good constraints
($L=10^{43.2-46}\text{ erg/s}$), the obscured fraction rises with
redshift so that at $z>2.25$, $\valuetotalobscspacedensitypeakzfrac$
of AGN are obscured ($N_{H}>10^{22}\text{ cm}^{-2}$), as compared
to the local $z=0$ value $z=\valuetotalobscspacedensitylowzfrac$. 

Our analysis also establishes a decline of the obscured AGN fraction
at low luminosities. This trend has only been found in the local Universe
\citep{Burlon2011,Brightman2011b}. Figure \ref{fig:obscured-fraction-lum}
also leads to the conclusion that observers who would consider only
intrinsically faint AGN ($L\leq10^{43}\text{ erg/s}$) would observe
a decrease in the fraction of obscured AGN with redshift. In other
words, the magnitude of the trend is determined by the sample selection.

\begin{figure}[h]
\begin{centering}
\includegraphics[bb=0bp 100bp 342bp 335bp,clip,width=0.5\textwidth]{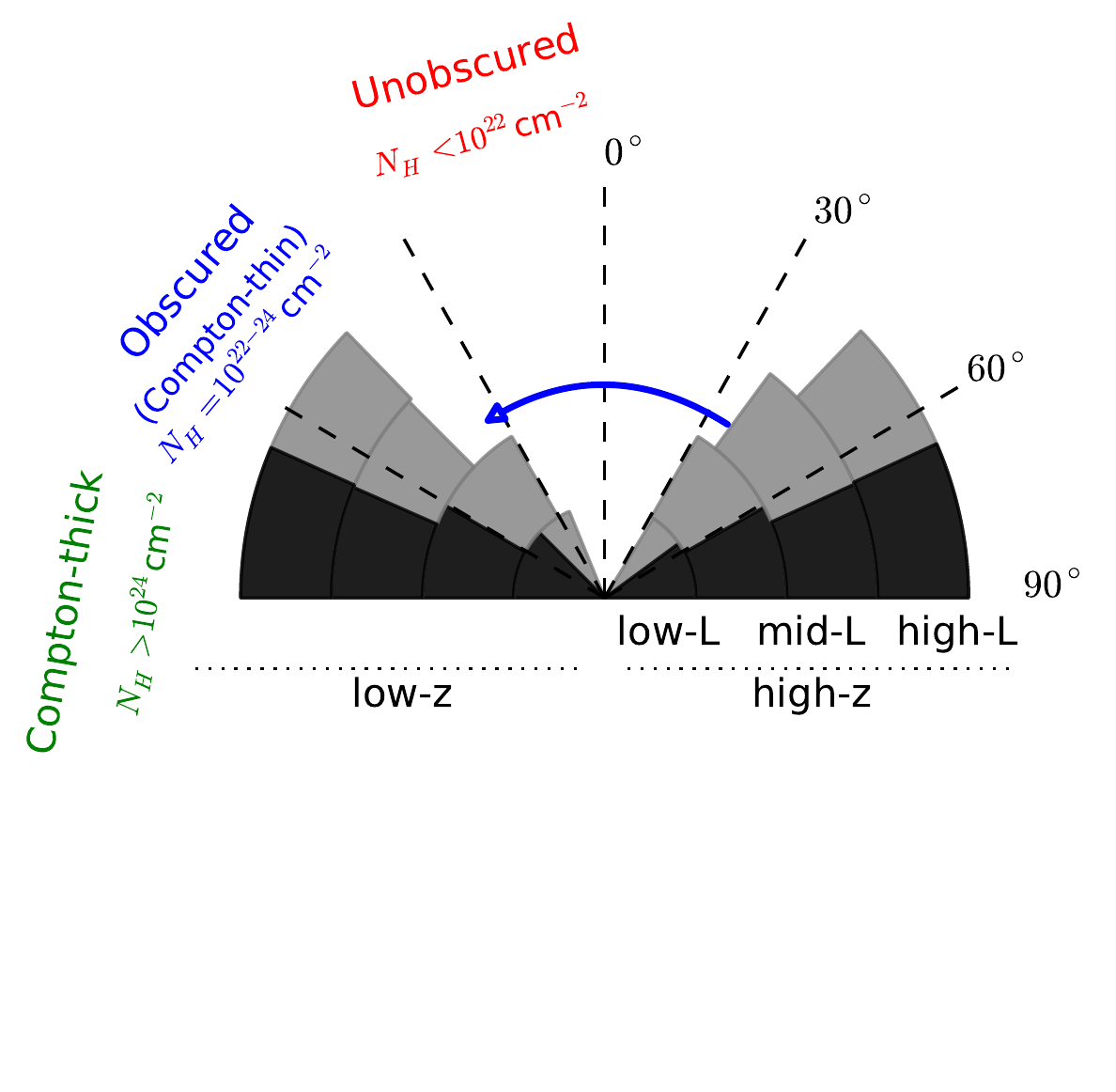}
\par\end{centering}

\protect\caption{\label{fig:Cartoon-obscfrac}Illustration of the population-averaged
obscuring toroid. The obscured and Compton-thick fraction and their
luminosity dependence is shown using a corresponding opening angle
above which the sky is obscured to the X-ray emitter (center). The
luminosity dependence is depicted for low-redshift (left) and high-redshift
(right) AGN using radial shells: as the luminosity increases, more
sky is visible. Notice that the difference between a low-redshift
and high-redshift source does not affect the lowest and highest luminosities,
but only where the transition occurs (blue arrow). The opening angle
corresponding to the Compton-thick transition remains roughly constant.
For accurate numbers, refer to Figures \ref{fig:obscured-fraction-lum}
and \ref{fig:obscured-fraction-evolution}.}
\end{figure}

\begin{figure*}[t]
\begin{centering}
\includegraphics[angle=90,height=9cm]{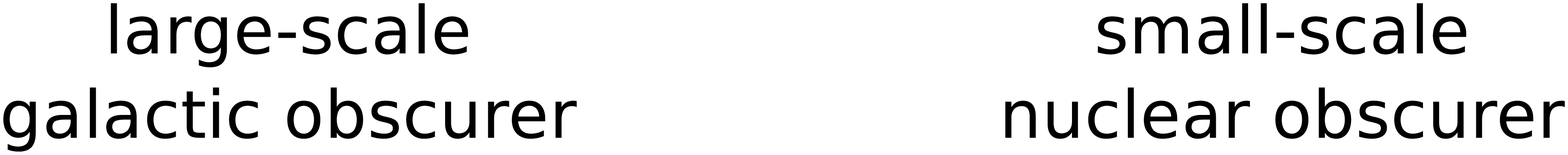}\includegraphics[height=9cm]{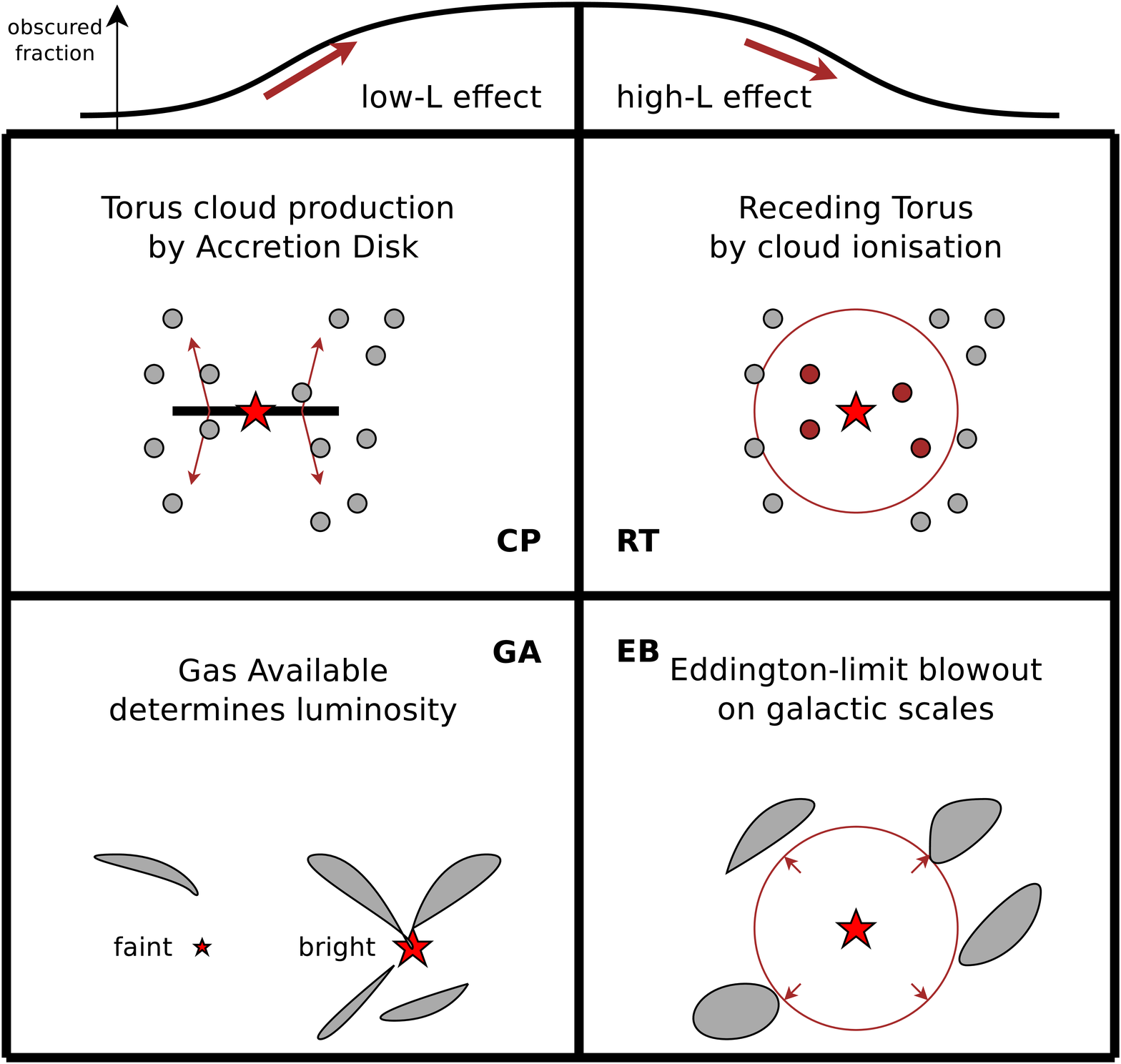}
\par\end{centering}

\protect\caption{\label{fig:Models}Four models that address the luminosity-dependence
of the obscured fraction. This categorisation distinguishes by the
luminosity-dependent effect, the increase of the obscured fraction
(left column) and the decrease of the obscured fraction (right column).
The top of Figure \ref{fig:obscured-fraction-lum} is repeated here
and simplified. In the four models considered (see text for details),
the relevant obscurer is either the torus (top row) or galactic obscuration
(bottom row). In the illustration of the models, the red star represents
the X-ray source, while grey clumps illustrate the cold obscurer. }
\end{figure*}

Obscurer models that do not invoke any obscuration-dependence are
ruled out by our findings. These include the simple torus where the
obscuration distribution is produced purely geometrically. Despite
these shortcomings, this picture may serve as a visualisation of the
observed trends. The luminosity-dependence and evolution of obscured/Compton-thick
AGN is illustrated in Figure \ref{fig:Cartoon-obscfrac}. If we assume
a toroidal geometry for the obscurer shared by the whole population,
we can define an opening angle that reproduces the obscured/Compton-thick
fraction. The luminosity dependent increase of this opening angle
from $30\text{\textdegree}$ to $40\text{\textdegree}$ is depicted
using radial shells for low-redshift (left) and high-redshift (right)
AGN. Notice that the difference between a low-redshift and high-redshift
source does not affect the lowest and highest luminosities, but only
where the transition occurs (blue arrow). Thus, it is not the case
that high-redshift sources merely have more obscuration overall. If
this were the case, a boost in the obscured fraction would be visible
over all luminosities. Instead the peak of the obscured fraction has
moved to lower luminosities, thus reducing the obscurer in moderately
bright AGN (see Figure \ref{fig:obscured-fraction-lum} for accurate
numbers).

We now consider a few exemplary models that can qualitatively account
for the decrease of the obscured fraction towards low luminosity or
the decrease towards high luminosities. An additional uncertainty
is whether the relevant obscuration occurs on small or large scales,
i.e. in the torus or on galactic scales. In the following, we present
for both scenarios a model each for the high-luminosity\change{,}{and}
the low-luminosity effect. These are illustrated in Figure \ref{fig:Models}.

The receding torus (RT) model is often invoked to explain the observed
decrease of the obscured AGN fraction at bright luminosities. This
scenario postulates a toroidal geometry for the obscurer with opening
angle that increases with increasing luminosity. This is attributed
to photon pressure pushing away the obscuring material, photo-ionisation
of the gas clouds or sublimation of the dust by the photon field of
the active black hole \citep{Lawrence1991,Simpson2005,Akylas2008}.
\change{}{Such a variation of the scale height of the torus has been
claimed via infrared observations \citep{Maiolino1995,Lusso2013,Toba2014}}.
In the context of clumpy torus models \citep{Hoenig2007,Nenkova2008},
a more intense radiation field leads to \change{the }{}the (partial)
ionisation of individual dust and gas clouds thereby increasing the
effective solid angle of unobscured sight-lines. This model (RT) thus
represents a direct, causal connection between the X-ray luminosity
and the unobscured line of sight. This scenario is illustrated in
the top right of Figure \ref{fig:Models}. In the RT model it is,
however, difficult to explain the observed evolution of the luminosity
dependence. The main evolutionary effect observed is an increase of
the turnover luminosity with redshift, causing the onset of the high-luminosity
effect at subsequently higher luminosities (see Figure \ref{fig:obscured-fraction-lum}).
For a purely nuclear obscurer scenario such as RT, no such evolutionary
effect is predicted.

Alternatively, it is also possible to interpret the observed luminosity
dependence of the obscured AGN fraction in the context of evolutionary
models, in which different levels of obscuration roughly correspond
to different stages of the growth of supermassive black holes. Most
black hole formation models include an early stage of gas inflows
caused by mergers \citep[e.g.][]{Hopkins2006OriginModel,Somerville2008}
or secular processes \citep[e.g.][]{Fanidakis2012,Bournaud2007,Ciotti_Ostriker1997,Ciotti_Ostriker2001}.
During this period super-massive black holes grow fast by accreting
material close to the Eddington limit and are also typically obscured
by the inflowing material. Eventually however, the energy output of
the central engine becomes sufficiently powerful to drive outflows,
which can blow away the obscuring clouds of dust and gas. The central
engine then shines unobscured for a brief period before its luminosity
output declines as a result of the depletion of the available gas
reservoirs. It is further proposed \citep[e.g.][]{Hopkins2005LFinterpretation,Hopkins2006Model}
that luminous AGN brighter than the break of the X-ray luminosity
function are dominated by fast accreting black holes close to the
peak of their growth phase. Lower luminosity sources include a large
fraction of AGN during the decline stage of their activity. 

In the above scenario the decrease of the obscured AGN fraction with
increasing luminosity may be linked to the blow-out stage of AGN.
This scenario (EB) illustrated in the bottom right panel of Figure
\ref{fig:Models}. At brighter accretion luminosities it is more likely
that AGN-related feedback mechanisms become more efficient, thereby
sheding the gas and dust cocoons in the vicinity of supermassive black
holes. Lower accretion luminosities, below the break of the X-ray
luminosity function, include a large fraction of AGN past the peak
of their activity, when the gas reservoirs that feed and obscure the
black hole have already been depleted. Therefore, one might expect
a larger fraction of unobscured AGN at low accretion luminosities.
The evolutionary picture outlined above is therefore consistent with
the observed luminosity dependence of the obscured AGN fraction: a
maximum at a given luminosity and a decline at both brighter and fainter
luminosities. 

In the EB model, the observed redshift dependence of the turnover
luminosity is also difficult to explain. Unless the physical conditions
of the obscuring material are a function of cosmic time, it is not
obvious how the physical mechanisms that affect the geometry of the
obscurer on small-scales can produce a luminosity dependent obscured
AGN fraction that changes with redshift. 

A rescuing argument for the EB scenario is possible. It is now fairly
established that the black hole mass of the average, currently accreting
AGN increases with redshift \citep{Merloni2004,Merloni2011}. This
means that at high redshift, the typical black hole undergoing accretion
has already accumulated a large mass, while at low redshift, small
black holes undergo accretion. Recent simulations have shown that
such a anti-hierarchical growth is not contradictory to hierarchical
structure formation \citep[e.g.][]{Fanidakis2012,Hirschmann2012,Enoki2014},
and can explain the decrease of the peak of the luminosity function.
Under the EB scenario, the peak of the luminosity function is linked
to the Eddington luminosity, as brighter AGN remove their obscuration.
If the black hole mass increases with redshift, so does the Eddington
luminosity, causing the on-set of the high-luminosity decrease of
the obscured fraction to occur at higher luminosities. The EB scenario
is thus compatible with our observations, if the average black hole
mass increases with redshift.

This model can affect both the small-scale, nuclear obscurer as well
as the gas in the galactic vicinity of the black hole, and our observations
may be incapable of distinguishing the effects. For this reason, the
illustration of this effect (EB) in Figure \ref{fig:Models} might
be placed in both categories.

Lets now consider the low-luminosity effect, which is a\change{n}{}
decrease of the obscured fraction towards low luminosities. The evolutionary
scenario might also be capable of explaining the low luminosity increase
of the obscuration \citep{Hopkins2006OriginModel}. If the observed
column density is due to galactic streams, feeding the black hole
simultaneously obscures it. One would then expect a simple correlation
of the accretion luminosity and the obscuration, at least at luminosities
below the Eddington limit. This effect (Gas Available, GA) is illustrated
in the bottom left of Figure \ref{fig:Models}.

Under the GA scenario (bottom left of Figure \ref{fig:Models}), the
accretion luminosity is simply correlated to the available gas. While
observations have shown an overall reduction of the gas and dust content
of galaxies over cosmic time \citep{Tacconi2013}, under the GA model
this would increase the total number of AGN at high luminosities,
but the luminosity-dependence of the obscured fraction would remain
unaffected. In particular, the GA model does not predict the observed
increase of the turn-over luminosity as suggested by Figure \ref{fig:obscured-fraction-lum}. 

The low-luminosity increase of the obscured fraction has been observed
before in local samples \citep{Burlon2011,Brightman2011b,Elitzur2009}.
For these sources, it is more probable that the obscurer is nuclear,
as many of these sources provide the opportunity to study both the
galaxy and the infrared emission of the torus. One model is that the
obscurer is caused by clumps in the disk wind \citep{Elitzur2006}.
This scenario was originally created to explain the vertical support
for a cold dusty torus that would otherwise collapse \citep{Krolik1988}.
In the disk wind model, the accretion disk is only capable of projecting
clouds beyond a certain luminosity, predicting the absence of a clumpy
torus in low-luminosity AGN. This scenario of cloud production (CP)
is illustrated in the top left of Figure \ref{fig:Models}.

While the cloud production model reproduces the necessary increase
of the obscured fraction, it does not predict any evolution over cosmic
time when taken at face value. However, under this model, the critical
luminosity below which the obscurer can not be sustained is strictly
determined by the black hole mass \citep{Elitzur2006,Elitzur2008}.
This model has received observational support in \citet{Elitzur2009}
who observe a black hole mass dependence of the luminosity dependence
from a local survey. This model, taken together with the black hole
mass evolution explained above, can thus explain the observed evolution.

\begin{table}[h]
\protect\caption{\label{tab:prediction-model-summary}\changed{Predictions of the
discussed models.}}

\centering
\begin{threeparttable}

\begin{centering}
\begin{tabular}{c>{\centering}p{8em}>{\centering}p{2em}>{\centering}p{2em}>{\centering}p{7em}}
\toprule[1.5pt] Model\tnote{a} & Obscuration\tnote{b} & Low-L effect\tnote{c} & High-L effect\tnote{d} & Evolution\tnote{e}\tabularnewline
\midrule RT & nuclear & $\checkmark$ &  & BHM\tnote{f}\tabularnewline
CP & nuclear &  & $\checkmark$ & no\tabularnewline
EB & nuclear or galactic &  & $\checkmark$ & BHM\tnote{f}\tabularnewline
GA & nuclear & $\checkmark$ &  & no\tabularnewline
\midrule RT+CP & nuclear & $\checkmark$ & $\checkmark$ & not for high-L\tabularnewline
GA+EB & galactic & $\checkmark$ & $\checkmark$ & not for low-L\tabularnewline
CP+EB & nuclear or both & $\checkmark$ & $\checkmark$ & BHM\tnote{f}\tabularnewline
\bottomrule[1.5pt]
\end{tabular}
\par\end{centering}

\begin{tablenotes}

\item[a] Model name as presented in Figure \ref{fig:Models}.

\item[b] Whether the luminosity-dependent layer of Obscuration is
associated with galaxy-scale obscuration (``galactic'') or the ``torus''
(``nuclear'').

\item[c] Whether the model predicts a decrease of the fraction of
obscured AGN towards low luminosities.

\item[d] Whether the model predicts a decrease of the fraction of
obscured AGN towards \change{low}{high} luminosities.

\item[e] Whether the model predicts any evolution of the effect with
cosmic time.

\item[f] The model scales with black hole mass. Evolution is predicted
if the average accreting black hole mass changes over cosmic time.

\end{tablenotes}

\end{threeparttable}
\end{table}

To conclude, we have considered four effects (simple models) which
are summarized in Table \ref{tab:prediction-model-summary}. Individually,
they only partially explain the observations. However, the CP model,
taken together with the EB scenario can explain the observations if
the black hole mass function evolves with redshift. Under this view,
the obscurer is vertically extended above a certain luminosity, and
the torus disappears when the Eddington luminosity drives away the
obscurer, unveiling the bright X-ray source for a (cosmically) brief
time. Here it remains uncertain whether the obscuration removed is
galactic, nuclear, or both. Both effects engage at a luminosity that
is dependent on the black hole mass, which provides the cosmic evolution
effect.

\change{}{Future observations may shed light on the dependence of
the galactic gas with the AGN X-ray luminosity. For instance, ALMA
(Atacama Large Millimeter/Submillimeter Array, \citealp{Beasley2006ALMA})
measurements of gas fractions in host galaxies of unobscured and obscured
AGN can be compared. If they do not differ, the GA model is ruled
out. A similar approach by comparing luminous and faint AGN can also
probe the predictions of the EB model, and establish whether the obscuration
is nuclear. Under the EB model, gas motions are expected in bright
AGN, which may be detectable with future X-ray spectroscopes on the
ATHENA mission \citep{Nandra2013}. Furthermore, to establish that
the evolutionary trend is caused by black hole mass evolution, black
hole mass-matched samples of luminous AGN may be considered. If such
a sample shows no redshift evolution in the luminosity-dependence
of the obscured fraction, then the trend is due to black hole mass
differences. The CANDELS fields with their high-resolution near-infrared
imaging is the best candidate for such a study. }

\change{}{The models as presented here are a rudimentary description
of the involved physical processes and require further refinement
in their predictions and self-consistency, e.g. via numerical simulations.
Among the challenges in this regard is that the nuclear obscuration
can not be resolved in evolutionary models \citep[e.g.][]{Hopkins2006OriginModel,Somerville2008}
and thus the torus is not yet treated self-consistently. Recently,
it has become possible to self-consistently model the radiative and
hydrodynamic processes that maintain the torus \citep[e.g.][]{Wada2012}.
Further research is needed in terms of whether the torus can, by itself,
reproduce the luminosity dependence of the obscured fraction. Also,
it has to be clarified which processes maintain a cold, Compton-thick
torus, with the geometric extent implied by a 35\% Compton-thick fraction.}

\section{\label{sec:Conclusions}Conclusions}

We combined deep and shallow, wide-area X-ray surveys conducted by
Chandra and XMM-Newton to constrain the space density of X-ray selected
AGN as a function of accretion luminosity, obscuring column density
and redshift. An important feature of our analysis is the non-parametric
method developed, which does not require any assumptions on the shape
of the luminosity function or its evolution, allowing the data to
drive our results. Furthermore, we take into account all sources of
uncertainties, allowing us to robustly constrain the evolution of
unobscured and obscured AGN, including the most heavily obscured Compton-thick.

We find that obscured AGN with $N_{H}>{\rm 10^{22}\,\text{ cm}^{-2}}$
account for $\valuetotalobscspacedensityaveragefrac$ of the number
density and $\valuetotalobsclumdensityaveragefrac$ of the luminosity
density of the accretion SMBH population averaged over cosmic time.
Compton-thick objects, with $N_{H}>{\rm 10^{24}\,\text{ cm}^{-2}}$
account for approximately half the number and luminosity density of
the obscured population, and $\valueCTspacedensityaveragefrac$ of
the total.

There is evidence that the space density of obscured, Compton-thin
AGN evolves stronger than the unobscured or Compton-thick AGN. This
is connected to the luminosity-dependent fraction of obscured AGN.
At higher luminosities, fewer AGN are obscured. However, at higher
redshift, this effect sets on at significantly higher luminosities.
In the luminosity range used in this study, the fraction of obscured
AGN increases from $\valuetotalobscspacedensitylowzfrac$ to $\valuetotalobscspacedensitypeakzfrac$
at $z>2.25$. This is due to only a small luminosity range around
$L_{*}\thickapprox10^{44}\,\text{ erg/s}$ which changes its obscured
fraction. In particular, the space density evolves fastest around
$N_{H}\approx10^{23}\text{ cm}^{-2}$. In contrast the fraction of
Compton-thick AGN relative to the total population is consistent with
being constant at $\approx35\%$ independent of redshift and accretion
luminosity. The contribution to the luminosity density by Compton-thick
AGN is $\valueCTlumdensityaveragefrac$. The robust determination
of a large fraction of Compton-thick AGN consolidates AGN X-ray surveys
and studies of the cosmic X-ray background. It also implies a physically
extended nuclear obscurer.

We also find that the fraction of moderately obscured AGN decreases
not only to high but also to low luminosities, in qualitative agreement
with findings of local surveys. We have discussed the observed trends
of the obscured fraction with $L_{X}$ and $z$ in the context of
models that either assign obscuration to the torus or galaxy-AGN co-evolution
effects, with varying luminosity-dependent effects. Both classes of
models can qualitatively explain our results but require that SMBH
evolve in a downsizing manner, i.e. larger black holes form earlier
in the Universe.

\acknowledgments{

\paragraph{Acknowledgements: }

JB acknowledges financial support through a Max Planck society stipend.
We thank the builders and operators of \textit{Chandra X-ray Observatory}
and \emph{XMM-Newton}. This research has made use of software provided
by the Chandra X-ray Center (CXC) in the application package CIAO
and Sherpa. Additionally, the CosmoloPy (\url{http://roban.github.com/CosmoloPy/}),
BXA (\url{https://johannesbuchner.github.io/BXA/}) and pymultinest
(\url{https://johannesbuchner.github.io/PyMultiNest/}) software packages
were used.

AGE acknowledges the \textsc{thales} project 383549, which is jointly
funded by the European Union and the Greek Government in the framework
of the programme ``Education and Lifelong Learning''.

Funding for SDSS-III has been provided by the Alfred P. Sloan Foundation,
the Participating Institutions, the National Science Foundation, and
the U.S. Department of Energy Office of Science. The SDSS-III web
site is \url{http://www.sdss3.org/}. SDSS-III is managed by the Astrophysical
Research Consortium for the Participating Institutions of the SDSS-III
Collaboration.}

\bibliographystyle{apj}
\bibliography{agn,apj-jour}

\begin{appendix}

\section{\label{sub:LF-intro}Luminosity function analysis}

We here go through a derivation of the statistical footing of analysing
population demographics by reviewing and combining the works of \citet{2004AIPC..735..195L}
and \citet{Kelly2008}. Their main difference is whether the Binomial
distribution or its approximation, the Poisson distribution, is used.

We intend to estimate the number density per comoving volume (Mpc$^{3}$)
of AGN, as a function of various properties, specifically X-ray luminosity,
redshift and obscuring column density. This extended luminosity function
will describe the evolution of the X-ray population and all sub-populations
(e.g. Compton-thick AGN). The difficulty is that each of these properties
influences our ability to detect objects with such properties. Lets
assume we can describe the probability to detect a source with properties
$\mathcal{C}$ as $p(\mathcal{D},\mathcal{C})$.

After we analysed each object in detail, we can bin our sample in
so small bins that only $1$ item at most can be in each bin $\mathcal{C}_{i}$.
Then just applying the Binomial/Poisson distribution, assuming that
in $k$ bins a source is detected (and no detections in the other
$n-k$ bins), yields:

\begin{eqnarray}
B(k;n,p) & = & \underbrace{{n \choose k}\times\prod_{i=1}^{k}p(\mathcal{D},d_{i},\mathcal{C}_{i})}_{\square}\times\label{eq:basic-binomial-survey}\\
 &  & \underbrace{\prod_{i=k}^{n}p(\bar{\mathcal{D}},d_{i},\mathcal{C}_{i})}_{*}\nonumber 
\end{eqnarray}

\begin{eqnarray}
P(k;n,p) & = & \underbrace{\frac{1}{k!}\times\prod_{i=1}^{k}n\times p(\mathcal{D},d_{i},\mathcal{C}_{i})}_{\square}\times\label{eq:basic-poisson-survey}\\
 &  & \underbrace{\exp\left\{ -\sum_{i=1}^{n}p(\mathcal{D},d_{i},\mathcal{C}_{i})\right\} }_{*}\nonumber 
\end{eqnarray}

The first equation gives the likelihood based on the Binomial distributions,
and here $p(\bar{\mathcal{D}},d_{i},\mathcal{C}_{i})$ denotes the
probability of not detecting a source with properties $C_{i}$ and
having obtained the observed data $d_{i}$. In the Poisson formula,
only $p(\mathcal{D},d_{i},\mathcal{C}_{i})$ occurs, which instead
denotes the probability of detection.

We notice that the term marked with a star in Equation \ref{eq:basic-poisson-survey}
remains the same, regardless of the number of detections $k$ in the
sample (only dependent on $n$). The sum in this term therefore has
to be always the same, independent of the specific sample $d_{i}$.
In the Poisson formalism, the exponent of this term has a specific
meaning -- it refers to the expectation value of the sample (a a-priori
assumption). We can thus remove the dependency of the data and treat
it as an a-priori probability. Mathematically, we integrate over all
possible data $d_{i}$.

\begin{eqnarray}
P(k;n,p) & = & \exp\left\{ -\sum_{i=1}^{n}p(\mathcal{D}_{i},\mathcal{C}_{i})\right\} \times\\
 &  & \prod_{i=1}^{k}\left(n\times p(\mathcal{D}_{i},d_{i},\mathcal{C}_{i})\right)
\end{eqnarray}

We used $\sum_{i=1}^{n}p(\mathcal{D}_{i},d_{i},\mathcal{C}_{i})=\sum_{i=1}^{n}p(\mathcal{D}_{i},\mathcal{C}_{i})$.
By analogy, for the binomial distribution, we use $\int p(\bar{\mathcal{D}}_{i},d_{i},\mathcal{C}_{i})\, dd_{i}=p(\bar{\mathcal{D}}_{i},\mathcal{C}_{i})$.
Mathematically, we again integrate over all possible data $d_{i}$,
but for all non-detected sources (see \citet{Kelly2008}).

\begin{eqnarray}
B(k;n,p) & = & {n \choose k}\times\prod_{i=k}^{n}p(\bar{\mathcal{D}}_{i},\mathcal{C}_{i})\times\\
 &  & \prod_{i=1}^{k}p(\mathcal{D}_{i},d_{i},\mathcal{C}_{i})\\
 & = & {n \choose k}\times\exp\left\{ \sum_{i=k}^{n}\ln p(\bar{\mathcal{D}}_{i},\mathcal{C}_{i})+\right.\\
 &  & \left.\sum_{i=1}^{k}\ln p(\mathcal{D}_{i},d_{i},\mathcal{C}_{i})\right\} 
\end{eqnarray}

Now we replace our $n$ discrete bins by a continuum. The probability
$p(\mathcal{D}_{i},d_{i},\mathcal{C}_{i})$ can then be non-zero over
a range of the parameter space. After all, we do not know the true
parameter ${\cal C}$. Thus, we replace $\sum_{i}p(\mathcal{D}_{i},d_{i},\mathcal{C}_{i})$
by $\int_{\theta}p(\mathcal{D}_{i},d_{i}|\mathcal{C})\, d\mathcal{C}$.

\begin{eqnarray}
P(k;n,p) & = & \frac{1}{k!}\times\exp\left\{ -\int p(\mathcal{D},\mathcal{C})\, d\mathcal{C}+\right.\\
 &  & \left.\sum_{i=1}^{k}\ln\int p(\mathcal{D}_{i},d_{i},\mathcal{C})\, d\mathcal{C}\right\} 
\end{eqnarray}

\begin{eqnarray}
B(k;n,p) & = & {n \choose k}\times\exp\left\{ \right.\\
 &  & (n-k)\cdot\ln\int p(\bar{\mathcal{D}},\mathcal{C})\, d\mathcal{C}+\\
 &  & \left.\sum_{i=1}^{k}\ln\int_{\theta}p(\mathcal{D}_{i},d_{i},\mathcal{C})\, d\mathcal{C}\right\} 
\end{eqnarray}

We will now expand $p(\mathcal{D},\mathcal{C})$ and $p(\mathcal{D}_{i},d_{i},\mathcal{C})$
using conditional probabilities and discuss the meaning of the occuring
terms.

\begin{eqnarray}
p(\mathcal{D},\mathcal{C}) & = & p(\mathcal{C})\cdot p(\mathcal{D}|\mathcal{C})\label{eq:apriori-prob}\\
p(\mathcal{D},d_{i},\mathcal{C}) & = & p(\mathcal{C})\cdot p(d_{i}|\mathcal{C})\cdot p(\mathcal{D}|d_{i},\mathcal{C})\label{eq:data-prob}
\end{eqnarray}

Here, $p(\mathcal{C})$ is the probability of finding an object with
characteristics $\mathcal{C}$ ($\{\log L,z,\log N_{H}\}$ here).
This is essentially the luminosity function we are interested in.
The term $p(\mathcal{D}|\mathcal{C})$ in Equation \ref{eq:apriori-prob}
denotes the sensitivity to such objects. 

In Equation \ref{eq:data-prob}, the second term, $p(d_{i}|\mathcal{C})$,
is related to the spectral analysis. It denotes the likelihood that
this data was generated from a source with e.g. luminosity $L$ and
column density $N_{H}$. Finally, we have to consider the probability
of detecting this object, given the data and characteristics $\mathcal{C}$.
This is almost surely~$1$, as having data associated with an object
implies having detected it. However, there is a subtlety here that
has been overlooked so far. Some astronomers use the sensitivity $p(\mathcal{D}|\mathcal{C})$
-- specifically the area curve of where the object was detected --
here, in place of $p(\mathcal{D}|d_{i},\mathcal{C})=1$. But \citet{2004AIPC..735..195L}
makes a strong point arguing that the sensitivity function must not
be used here. Why is it done in practise anyways?

The X-ray source detection algorithms typically use a area of a certain,
relatively small size, which contains 70\% of the energy contributed
by the source. The background for this region is estimated, and the
probability for the background to produce the observed number of counts
computed (no-source probability). If a threshold is exceeded, the
spectrum is extracted. In this step a much larger region is used,
containing more background counts. While in the small detection area
it was not possible for the background to produce the observed counts,
for faint sources it may be possible for the background to produce
the observed counts in the spectrum extraction region. This is because
the background contribution grows linearly with area, but the source
contribution almost stays the same. The likelihood of our analysis
thus does allow having no flux from a faint source, i.e. luminosity
zero. This contradiction to the detection probability stems from the
fact that we have not used the information that the counts are \emph{concentrated}
in detection region. In this case, $p(\mathcal{D}|d_{i},\mathcal{C})\neq1$.

Specifically, we could write it as $p(\mathcal{D}|d_{i},\mathcal{C})=p(>k|d_{i},\mathcal{C})$,
the probability of having more than $k$ counts in the detection region
out of the extracted spectrum counts $d_{i}$, where $k$ denotes
the number necessary for detection at the current position. An approximation
to this number is $p(>k|\mathcal{C})$, the probability to produce
the number of counts required for a detection at the current position.
A further approximation is $\int p(>k|\mathcal{C})\, dA/A=p(\mathcal{D}|\mathcal{C})$,
which is the area-average sensitivity curve. Given this, we understand
now why some works use the area curve in the data term, even though
it should not be done: it is an attempt in fixing a loss of information
introduced when the detection and data analysis processes differ.

We now put all the information together and arrive at the relevant
likelihoods:

\begin{eqnarray}
B(k;n,p) & = & {n \choose k}\times\left(\int p(\bar{\mathcal{D}}|\mathcal{C})\cdot p(\mathcal{C})\, d\mathcal{C}\right)^{n-k}\times\label{eq:Kelly-eq-5}\\
 &  & \prod_{i=1}^{k}\int p(\mathcal{C})\cdot p(d_{i}|\mathcal{C})\cdot p(\mathcal{D}|d_{i},\mathcal{C}))\, d\mathcal{C}\nonumber 
\end{eqnarray}

\begin{eqnarray}
P(k;n,p) & = & \frac{1}{k!}\times\exp\left\{ -\int p(\mathcal{C})\cdot p(\mathcal{D}|\mathcal{C})\, d\mathcal{C}\right\} \times\label{eq:Loredo-eq-9}\\
 &  & \prod_{i=1}^{k}\int p(\mathcal{C})\cdot p(d_{i}|\mathcal{C})\cdot p(\mathcal{D}|d_{i},\mathcal{C}))\, d\mathcal{C}\nonumber 
\end{eqnarray}

Comparing Equation 9 in \citet{2004AIPC..735..195L} with Equation
\ref{eq:Loredo-eq-9} above, we have obtained the same result with
the notation key $p(d_{i}|\theta)=l_{i}$, $p(\mathcal{D}|d_{i},\mathcal{C})=1$,
$\mathcal{C}=m$, $n=\omega\delta m$ and $p(\mathcal{D}|\mathcal{C})=\eta(m)$.
The derivation is similar, however we started from first principles
and continued in a sound and complete way for both the Poisson and
the Binomial distribution.

Comparing Equation 5 in \citet{Kelly2008} with Equation \ref{eq:Kelly-eq-5}
above, we have obtained the same result with the notation key ${\cal C}=(L_{j},z_{j})$,
$p(\bar{\mathcal{D}}|\mathcal{C})=p(I_{j}=0|L_{j},z_{j})$, $p({\cal C})=p(L_{j},z_{j}|\theta)$,
$p(\mathcal{D}|d_{i},\mathcal{C})=1$ as well as $p(d_{i}|\mathcal{C})=1$,
since \citet{Kelly2008} neglects any measurements uncertainties in
the derivation, assuming that luminosity and redshift can be determined
perfectly.

Finally, we summarise the logarithms of the likelihoods, neglecting
constants.

\begin{eqnarray}
\ln{\cal L}_{B} & = & (n-k)\times\ln\int p(\bar{\mathcal{D}}|\mathcal{C})\cdot p(\mathcal{C})\, d\mathcal{C}+\label{eq:Binom-like}\\
 &  & \sum_{i=1}^{k}\ln\int p(\mathcal{C})\cdot p(d_{i}|\mathcal{C})\cdot p(\mathcal{D}|d_{i},\mathcal{C})\, d\mathcal{C}\nonumber 
\end{eqnarray}
\begin{eqnarray}
\ln{\cal L}_{P} & = & -\int p(\mathcal{C})\cdot p(\mathcal{D}|\mathcal{C})\, d\mathcal{C}+\label{eq:Poisson-like}\\
 &  & \sum_{i=1}^{k}\ln\int p(\mathcal{C})\cdot p(d_{i}|\mathcal{C})\cdot p(\mathcal{D}|d_{i},\mathcal{C})\, d\mathcal{C}\nonumber 
\end{eqnarray}

The general-purpose framework presented here is advocated by \citet{Kelly2008}.
It is a general approach for the inference of population demographics
based on specific samples. The importance of this step -- rather than
analysing trends based on sample statistics -- and incorporating selection
biases, can not be over-stated.

In this work, we restrict ourselves to the Poisson likelihood. This
allows us to use the space density directly (rather than separating
sampled volume and probability). For $p(\mathcal{C})$ we insert the
luminosity function model $\phi(\log L,\, z,\,\log N_{H})\times\frac{dV}{dz}$.
For $p(\mathcal{D}|\mathcal{C})$ we insert the area curve. For $p(\mathcal{D}|d_{i},\mathcal{C})$
we use the area curve of the specific field as an approximation (see
above). For $p(d_{i}|\mathcal{C})$ we should use the likelihood of
the spectral analysis. However, we did use intermediate priors in
the data analysis. We thus extend $p(d_{i}|\mathcal{C})=\frac{p(d_{i}|\mathcal{C})\times p(\mathcal{C})}{p(\mathcal{C})}$
where the top term is the posterior distribution computed in the spectral
analysis, and the prior has to be divided away again. As we used flat
priors in $\log L,\,\log N_{H}$, and $z$ (subject to additional
information from other wavelengths), which are the units of the integral
over \foreignlanguage{english}{$\mathcal{C}=\{\log L,\, z,\,\log N_{H}\}$},
this division only contributes to the likelihood as a fixed offset,
and is not relevant for further analysis. If different intermediate
priors had been used, here they would be have to be divided away.
In fact, we employed a Gaussian prior on the photon index $\Gamma$.
The luminosity functions $\phi$ we consider however are in fact $\phi(\log L,\, z,\,\log N_{H})\times f(\Gamma)$
where $f$ is the Gaussian prior used, thus we also do not require
a correction here. 

Regarding the practical evaluation of the likelihood function, we
use posterior samples from our spectral analysis as just described.
For the first integral, which describes the expected number of detected
sources, we employ a fixed grid, as these are fast to compute but
also ensure a consistent estimate between likelihood evaluations.
We use 40 grid points in each dimension. By pre-computation of all
weights except for the luminosity function $\phi$, the likelihood
function reduces to (fast) addition and multiplication operations.

\begin{algorithm*}

\begin{lstlisting}[basicstyle={\scriptsize\ttfamily},breaklines=true,numbers=left]
data {
  int indices[27344,3];
  real weights[27344];
  vector[11-1] widths0;
  vector[14-1] widths1;
  vector[7-1] widths2;
  int lengths[2046];
  int chain_indices[780,3];
  real chain_weights[780];
}
parameters {
  real<lower=-40,upper=0> y[11,14,7];
}
model {
  real sigmaL;
  real sigmaz;
  real sigmaNH;
  real dataterms[2046];
  real detectionterms[780];
  real loglike;

  sigmaL <- 0.5;
  sigmaz <- 0.5;
  sigmaNH <- 0.75;
  
  /* L smoothness prior: 2nd derivative is small */
  for (i in 3:11) {
    for (j in 1:14) {
      for (k in 1:7) {
        y[i,j,k] ~ normal((y[i-1,j,k] - y[i-2,j,k]) * widths0[i-1] / widths0[i-2] + y[i-1,j,k], sigmaL);
  } } }

  /* z smoothness prior: 2nd derivative is small */
  for (i in 1:11) {
    for (j in 3:14) {
      for (k in 1:7) {
        y[i,j,k] ~ normal((y[i,j-1,k] - y[i,j-2,k]) * widths1[j-1] / widths1[j-2] + y[i,j-1,k], sigmaz);
  } } }

  /* NH smoothness prior: 1st derivative is small */
  for (i in 1:11) {
    for (j in 1:14) {
      for (k in 2:7) {
        y[i,j,k] ~ normal(y[i,j,k-1], sigmaNH);
  } } }
  
  { /* individual objects */
  int m;
  m <- 0;
  for (i in 1:2046) {
    int a; int b; int c;
    real v[lengths[i]];
    
    for (j in 1:lengths[i]) {
      int l;
      l <- m + j;
      /* look up interpolation point and compute weights */
      a <- indices[l,1] + 1;
      b <- indices[l,2] + 1;
      c <- indices[l,3] + 1;
      v[j] <- exp(y[a,b,c]) * weights[l];
    }
    m <- m + lengths[i];
    /* add value to likelihood */
    dataterms[i] <- log(sum(v) / lengths[i]);
  }
  }
  
  /* detection integral */
  for (k in 1:780) {
    int a; int b; int c;
    real v;
    /* look up interpolation point and compute weights */
    a <- chain_indices[k,1] + 1;
    b <- chain_indices[k,2] + 1;
    c <- chain_indices[k,3] + 1;
    v <- chain_weights[k] * exp(y[a,b,c]);
    detectionterms[k] <- v;
  }
  loglike <- sum(dataterms) + sum(detectionterms);
  increment_log_prob(loglike);
}
\end{lstlisting}

\protect\caption{\label{alg:Stan-code}Stan code for estimating the field. The contant-slope
prior implementation is shown. We prepared the integrals to be computed
using importance sampling chain points evaluated the respective field
bin value (indices) and multiplied by the weights.}

\end{algorithm*}

\section{Robust photometric redshift probability distributions\label{sec:pdzsys}}

In this section we elaborate on incorporating systematic uncertainties
into photometric probability distributions from SED fitting.

The photometric fitting method of \citet{Salvato2011,Salvato2009},
using software from \citet{Ilbert2006,Ilbert2009} is capable of not
just computing a most likely redshift point estimate, but producing
redshift probability distributions. Previous studies, such as \citet{Buchner2014},
used the photo-z probability distributions directly to incorporate
the uncertainty of the redshift estimate. However, additional to the
uncertainty due to measurement errors, the method of photometric redshift
has systematic errors due to incomplete template libraries, failures
in automatically extracting correct fluxes at the edge of images,
blending of sources causing the flux to be convoluted, etc. The most
important systematic contribution is probably incorrect associations.

On a high-level, the quality of redshift point estimates for a specific
sample can be described by two quantities, which are estimated from
a sub-sample where spectroscopic redshifts are available: the outlier
fraction $\eta$ and the scatter $\sigma$ (\citealp{Salvato2011}).
The outlier fraction describes the failure rate or accuracy of the
redshifts, i.e. how frequent the redshift estimate is off by far:

\[
\eta:\,\text{fraction where}\left|\frac{z_{\text{phot}}\text{\textminus}z_{\text{spec}}}{1+z_{\text{spec}}}\right|>0.15
\]
The scatter width $\sigma$ describes how precise the redshifts are
scattered around the true value. The sample can be characterised using
the normalised median absolute deviation \citep[NMAD,][]{Hoaglin1983NMAD}
as a robust estimator:

\[
\sigma_{\text{NMAD}}=1.48\times\text{median}\left|\frac{z_{\text{phot}}\text{\textminus}z_{\text{spec}}}{1+z_{\text{spec}}}\right|
\]

These two quantities are interpreted that the redshift estimates $z_{\text{phot}}$
are distributed along a normal distribution centred at the true value
$z_{\text{spec}}$ with standard deviation $\sigma_{\text{NMAD}}\cdot\left(1+z_{\text{spec}}\right)$.
Additionally, a fraction $\eta$ of the redshift estimates are drawn
from a different distribution, which can be described as flat over
a redshift range $(0,\, z_{\text{max}})$. Here we considered $z_{\text{max}}=5$
for outliers. The probability distribution of the point estimator
incorporating systematic errors (SYSPE) can be written as

\[
\text{SYSPE}(z):=\eta\times\text{U}(0,\, z_{\text{max}})+(1-\eta)\times\text{N}(z_{\text{phot}},\,\sigma)
\]
This modelling in fact matches the empirical distribution very well
when considering the cumulative distributions of $\left|z_{\text{phot}}\text{\textminus}z_{\text{spec}}\right|/\left(1+z_{\text{spec}}\right)$. 

In this work, we would like to similarly incorporate catastrophic
outliers. These are not contained in the photometric redshift distribution
(PDZ) from SED fitting. Thus, we follow an analoguous modeling of
the uncertainty contribution: We allow a broadening of the PDZ by
convolution with a Gaussian kernel of width $\bar{\sigma}$ and add
a flat probability plateau with weight $\bar{\eta}$:

\[
\text{SYSPDZ}(z):=\bar{\eta}\times\text{U}(0,\, z_{\text{max}})+(1-\bar{\eta})\times\left(\text{PDZ}*\text{N}(0,\,\bar{\sigma})\right)
\]
Note that for the special case $\bar{\sigma}=0$ and $\bar{\eta}=0$,
SYSPDZ becomes PDZ. The two parameters $\bar{\sigma}$ and $\bar{\eta}$
have a slightly different definition than $\sigma$ and $\eta$ above,
but take the same roles for characterise systematic uncertainty. We
fit for the parameters using the spectroscopic sub-sample by maximising
the likelihood

\[
{\cal L}(\bar{\sigma},\bar{\eta})=\prod_{i}\text{SYSPDZ}_{i}(z_{\text{spec},\, i})
\]
defined as the product of the value of the modified PDZ at the true
redshifts $z_{\text{spec}}$. PDZs that are far off the true value
will demand an increase in $\bar{\sigma}$ and $\bar{\eta}$, however
raising both values diminishes the SYSPDZ value overall due to normalisation
of probability distributions.

We find the best fit parameters $\bar{\sigma}=0.024,\,0.048,\,0.029$
and $\bar{\eta}=1.8\%,\,0.6\%,\,2.3\%$ for the CDFS, AEGIS-XD and
C-COSMOS samples respectively. These numbers are not directly equivalent
to the point estimate based definition of $\sigma_{\text{NMAD}}$
and $\eta$ (see \citealp{Salvato2011} and \citealp{HsuCDFSz2013}),
but the numbers are comparable. We thus use the smoothed SYSPDZ distributions
 instead of the PDZs.

A simpler method for incorporating the systematic uncertainty would
be to use the modelling based on the point estimator (SYSPE). But
when comparing the likelihood value, we find that SYSPE always has
lower likelihood values than the kernel smoothing SYSPDZ. This shows
that the PDZ contains valuable information due to e.g. secondary solutions,
which is missing in the point estimator $z_{\text{phot}}$, and that
SYSPDZ is the most faithful representation of the state of redshift
information.

\section{Luminosity function as table%
\footnote{Supplemental online material%
}\label{sec:LF-table}}

Available online at \url{http://www.mpe.mpg.de/~jbuchner/agn\_torus/paper/fieldsummary.cds}

\end{appendix}
\end{document}